\documentclass[12pt]{article}
\usepackage{amsmath}
\usepackage{graphicx,psfrag,epsf}
\usepackage{enumerate}
\usepackage[english]{babel}
\usepackage[utf8]{inputenc}
\usepackage{amsmath}
\usepackage{hyperref}
\usepackage{csquotes}
\usepackage{booktabs}
\usepackage{longtable}
\usepackage{array}
\usepackage{multirow}
\usepackage{wrapfig}
\usepackage{float}
\usepackage{colortbl}
\usepackage{pdflscape}
\usepackage{tabu}
\usepackage{threeparttable}
\usepackage{threeparttablex}
\usepackage[normalem]{ulem}
\usepackage{makecell}
\usepackage{xcolor}
\usepackage{natbib}
\usepackage{backref}
\renewcommand*{\backref}[1]{}
\renewcommand*{\backrefalt}[4]{
      \ifcase #1 (Not cited.) \ 
          \or [#2]        
          \else [#2]     
      \fi%
  }
\setcitestyle{notesep={;},round,aysep={},yysep={;}}
\newcommand{\blind}{0}

\begin{document}
\def\spacingset#1{\renewcommand{\baselinestretch}%
{#1}\small\normalsize} \spacingset{1}


\if0\blind
{
  \title{\bf Hierarchical Forecasting of Dengue Incidence in Sri Lanka}
  \author{L. S. Madushani\\
    Department of Mathematics, University of Sri Jayewardenepura, Sri Lanka\\
    and \\
    Thiyanga S. Talagala \thanks{
    Corresponding author}\\
    Department of Statistics, University of Sri Jayewardenepura, Sri Lanka}
  \maketitle
} \fi

\if1\blind
{
  \bigskip
  \bigskip
  \bigskip
  \begin{center}
    {\LARGE\bf Title}
\end{center}
  \medskip
} \fi

\bigskip
\begin{abstract}
The recurrent thread of dengue incidence in Sri Lanka is still abundant and it creates a huge burden to the country. Hence, the National Dengue Control Unit of Sri Lanka propose a national action plan to prevent and control the dengue incidence. To implement the necessary actions for short and long terms the proposed plan operates under three levels: country-level, province-level, and district levels. In order to optimize resource allocation, the health officers require the forecasts for country, province and district levels, which preserves the aggregate consistency associated with the district, province, and country levels as well as time correlations. Hence, the objective of this study is to forecast the dengue incidence in Sri Lanka using a hierarchical time series forecasting approach based on the spatial and temporal hierarchical structures. Hierarchical forecasting involves in two steps such as generating base forecasts and reconciliation of these base forecasts. In this study, Exponential smoothing (ETS) and Autoregressive Integrated Moving Average
(ARIMA), NAIVE, Seasonal NAIVE  approaches and average method are used to generate base forecasts. Accuracy of forecasts is evaluated using Mean Absolute Scale Error (MASE). We compare the accuracy of hierarchical time series forecasts with other benchmark approaches. The forecast accuracy reveals that the best forecasting approaches for country, provinces and districts are not limited to a single approach. Hence, we investigate reasons for the variations of performance in different forecasting approaches based on a time series feature-based visualization approach. 
\end{abstract}

\noindent%
{\it Keywords:}  Epidemiology, aggregate consistency, principal component analysis, time series features

\spacingset{1.45}
\section{Introduction}
\label{sec:intro}
Dengue is the most rapidly growing viral infection in the world \citep{world2014dengue}. The causative agent of dengue infection is classified into four serotypes: DENV-1, DENV-2, DENV-3, and DENV-4 whereas infection of one serotype does not result in lifelong immunity against other serotypes \citep{sanyaolu2017global}. According to the World Health Organization, 50 to 100 million new dengue cases occur annually including 20,000 deaths in more than 100 countries of the world \citep{world2012global}. A large number of dengue cases were reported in  Central and South America, South-East Asia and Western Pacific regions \citep{teixeira2009diagnosis}. Asia represents approximately 70$\%$ of the global burden of the disease \citep{bhatt2013global}.

Dengue incidence of Sri Lanka shows a considerable rise during the last two decades, causing economic and social burdens. The first serological confirmed dengue case was reported in 1962 and the first largest dengue epidemic outbreak occurred in 2009 with 35,008 suspected cases and 346 deaths in Sri Lanka \citep{withanage2018forecasting}. The largest dengue outbreak was reported in 2017, comprising 186,101 suspected cases and over 320 deaths according to the Epidemiological Unit of the Ministry of Health, Sri Lanka \citep{chandrakantha2019risk}. This unprecedented outbreak is due to the variation of the usual circulating DENV-2 \citep{ali2018unprecedented, tissera2020severe}. More than 30,000 cases have been reported each year since 2012 and the highest number of dengue cases is reported in the Western province of Sri Lanka \citep{withanage2018forecasting}. At present, the spread of dengue incidence in districts outside the Western province has indicated a dramatic change. 

To date, there is no specific treatment methodology to cure the infection \citep{wijegunawardana2019evaluation}. However, simple fluid replacement and case management assist to reduce the fatality rate from 20$\%$ to 1$\%$ \citep{chandrakantha2019risk}. Therefore, forecasting dengue incidence is important to implement curative and control programs. In recent, several researchers have made attempts to forecast dengue incidence in Sri Lanka. For example, time series regression models were developed using different climate variables for various districts \citep{withanage2018forecasting, goto2013analysis}. Distributed lag nonlinear modeling approach was used to identify the relationship between climate factors and dengue incidence in Colombo district of Sri Lanka \citep{talagala2015distributed}. \citet{talagala2015wavelet} performed a wavelet analysis to identify the pattern of dengue incidence in 25 districts of Sri Lanka . More recently, \citet{chandrakantha2019risk} developed a model to predict the likelihood of having dengue incidence in Sri Lanka based on climate factors. Furthermore, artificial neural network approach was applied to forecast dengue incidence in Kandy district of Sri Lanka for the period 2010 to 2012 by considering rainfall, humidity, temperature factors, and previous dengue cases \citep{nishanthi2014prediction}. \citet{attanayake2020exponential}  used exponential smoothing approach to forecast dengue counts in Colombo. All these studies have been conducted to forecast dengue incidence of the whole country, particular provinces, or districts independently. None of these approaches has focused to cater the needs that are proposed by the national action plan to implement dengue prevention and control programs \citep{unit2019ministry}. This plan has assigned officers for national, provincial, and district levels for achieving the proposed objectives of dengue control and prevention activities. Hence, this study mainly focuses on dengue incidence forecasting for districts, provinces, and the whole country to assist the decision-making process of assigned officers and plan and implement the programs at the district, province, and country level. 

The hierarchical time series forecasting has the ability to deal with time series that can be aggregated at different levels based on geography. This approach provides forecasts that are coherent across the levels of hierarchy. That is, forecasts at the aggregate level are equal to the sum of the relevant forecast at the desegregated level \citep{wickramasuriya2019optimal}. Moreover, it allows to capture of the correlation and interaction between the series at each level of hierarchy and it has the ability to capture different information based on each series. In addition, the most significant aspect of this approach is that it ensures aligned decision-making across the entire hierarchical structure \citep{athanasopoulos2020hierarchical}. Hence, this study considers hierarchical time series forecasting approach to obtain the forecasts of districts, provinces, and country levels of Sri Lanka. Furthermore, forecasts are needed for short-term and long-term periods. Hence, this study gives attention towards the time correlation as well. 

\section{Methodology and Data}

\subsection{Forecasting hierarchical time series}
\label{sec:meth}

\begin{figure}[h!]
\centering
  \includegraphics[scale=0.5]{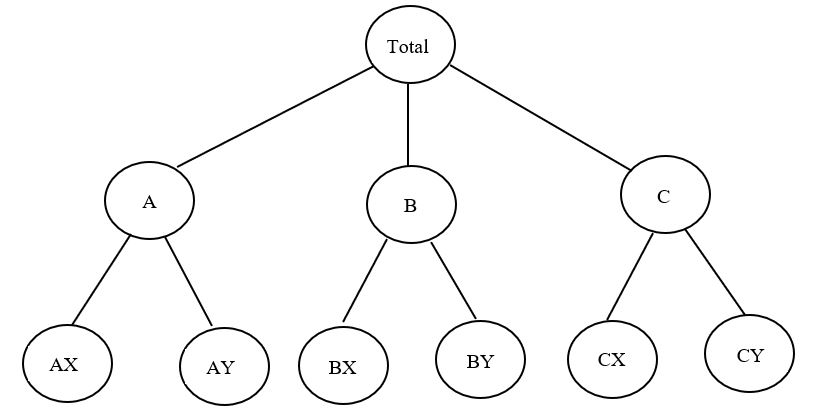}
  \caption{ Two-level hierarchical tree diagram}
  \label{fig:figurechapter21}
\end{figure}

A simple hierarchical structure is illustrated in Figure \ref{fig:figurechapter21}. In Figure \ref{fig:figurechapter21}, each series are observed at time $t=1,2,\dots,T$ and forecasts are generated for each series at time $T+1, T+2, \dots , T+h$.  The series at the bottom level of the hierarchy, level 2, are denoted as $AX, AY, BX, BY, CX$ and $CY$. Series at level 1, $A, B$ and $C$ can be obtained by aggregating the corresponding bottom level series. These series are further aggregated to obtain the total series at level 0, the "Total". Furthermore, $y_{i,t}$ denotes the $t^{th}$ series corresponds to node $i$ and $m_{i}$ denotes number of series at level $i$. The total number of series in hierarchical structure is denoted by $m$.
Hierarchical structure of Figure \ref{fig:figurechapter21} can be denoted by matrix notation as
\[
\left[\begin{array}
{r}
y_{Total,t}\\
y_{A,t}\\
y_{B,t}\\
y_{C,t}\\
y_{AX,t}\\ 
y_{AY,t}\\ 
y_{BX,t}\\ 
y_{BY,t}\\
y_{CX,t}\\
y_{CY,t}
\end{array}\right] = \left[\begin{array}
{rrrrrrrrr}
1 & 1 & 1 & 1 & 1 & 1\\ 
1 & 1 & 0 & 0 & 0 & 0\\ 
0 & 0 & 1 & 1 & 0 & 0\\ 
0 & 0 & 0 & 0 & 1 & 1\\
1 & 0 & 0 & 0 & 0 & 0\\
0 & 1 & 0 & 0 & 0 & 0\\  
0 & 0 & 1 & 0 & 0 & 0\\ 
0 & 0 & 0 & 1 & 0 & 0\\ 
0 & 0 & 0 & 0 & 1 & 0\\ 
0 & 0 & 0 & 0 & 0 & 1
\end{array}\right]
 \left[\begin{array}
{r}
y_{AX,t}\\ 
y_{AY,t}\\ 
y_{BX,t}\\ 
y_{BY,t}\\
y_{CX,t}\\
y_{CY,t}
\end{array}\right]
\]
In general, hierarchical time series can be denoted by the matrix notation $y_{t}=S b_{t}$, where $y_{t}$ denotes all series in the hierarchy and $b_{t}$ is the vector of all bottom level series in the hierarchy at time $t$. $S$ is the summing matrix of order $m \times m_{k}$ where $m_{k}$ is the number of series in the most disaggregate level of the hierarchical structure.

 Several methods have been proposed to reconcile the time
series forecasts in a way that respects the hierarchy.  However, all existing hierarchical time series forecasting approaches can be illustrated as $$\tilde{y}_{T}(h)=SP\hat{y}_{T}(h),$$ where $\tilde{y}_{T}(h)$ is $h$-step-ahead reconciled forecasts, $\hat{y}_{T}(h)$ is $h$-step-ahead base forecasts, and $P$ is an $m_{k}\times m$ matrix. The role of $P$ is different according to the hierarchical forecasting approach. 

Top-down and bottom-up approaches are the common approaches used in hierarchical time series forecasting. The bottom-up approach involves forecasting each of the bottom level series and sum these forecasts to obtain other levels of the hierarchy \citep{hyndman2011optimal}. However, bottom level data are noisy and therefore it is difficult to model bottom level series \citep{zotteri2014forecasting}. In the top-down approach, forecasts are generated for the most aggregated series and then forecasts of the total series are disaggregated based on the proportions. Here, the proportions can be obtained based on average or historical proportions \citep{athanasopoulos2009hierarchical,hyndman2011optimal}.

Another approach named as optimal combination approach  was proposed by \citep{hyndman2011optimal} which performs better than top-down and bottom-up approaches. This approach obtains the forecasts of all series of the hierarchical structure and then optimally reconcile based on a regression model. It requires an estimator for the covariance matrix of errors that occur due to incoherence. However, it is impossible to estimate. Hence, minimum trace optimal reconciliation approach (MinT) was introduced to incorporate the information from a full covariance of forecast errors by minimizing the mean squared error of the coherence forecasts under the assumption of unbiasedness and the resulting reconciled forecasts are in the form $$\tilde{y}_{T}(h)=S(S^{'}W_{h}^{-1}S)^{-1}S^{'}W_{h}^{-1}\hat{y}_{T}(h).$$ In practice, estimation of $W_{h}$ is difficult \citep{hyndman2018forecasting}. Hence, \citep{wickramasuriya2019optimal} proposed alternative estimators for $W_{h}$. 

The ordinary least square estimator (OLS) can be obtained with the substitution $$W_{h}=k_{h}I\hspace{0.2in}\forall h,$$ where $k_{h}>0$ and $I$ is an identity matrix.  The drawback of this substitution is that the scale differences between the levels of the structure are not considered. Another estimation is $$W_{h}=k_{h}diag(\hat{W}_{1})\hspace{0.2in}\forall h,$$ where $k_{h}>0$ and $\hat{W}_{1}=\frac{1}{T}\sum_{t=1}^{T}e_{t}{e_{t}}^{'}$. In this case, $e_{t}$ is in-sample residuals of the base forecasts stacked in the same order as the data. Here, the base forecasts are scaled using the variance of residuals. Therefore, this approach is referred to as MinT approach based on a weighted least square (WLS) estimator using variance scaling. Another criterion considers that each of the bottom-level base forecasts are uncorrelated between nodes and they have errors with equal variance $k_{h}$. Furthermore, each element of the diagonal matrix $\Lambda$ denotes the number of forecast error variances that contribute to that aggregation level. This estimator focuses only on the structure of the hierarchy as $$W_{h}=k_{h}\Lambda\hspace{0.2in}\forall h,$$ where $k_{h}>0$ and $\Lambda=diag(S1)$ with 1 being a unit vector of dimension $m_{k}$. Next, $$W_{h}=k_{h}W_{1}\hspace{0.2in}\forall h,$$ where $k_{h}>0$ is considered as the estimation. In this estimation, $W_{1}$ is full one-step covariance matrix. However, this estimator is not an appropriate estimator when the number of bottom level series is larger than the length of the series $T$. Hence, shrinkage estimator is provided based on the sample covariance to a diagonal matrix as $$W_{h}=k_{h}\hat{W}_{1,D}^{*}\hspace{0.2in}\forall h,$$ where $k_{h}>0$. In this method, $\hat{W}_{1,D}^{*}$ is a shrinkage estimator defined as $$\hat{W}_{1,D}^{*}=\lambda \hat{W}_{1,D}+(1-\lambda)\hat{W}_{1},$$ where $\hat{W}_{1,D}$ is a diagonal matrix that consists of diagonal entries of $\hat{W}_{1}$. Therefore, this estimator allows us to capture strong interrelations between time series in the hierarchy. 
\citet{schafer2005shrinkage} proposed the shrinkage intensity parameter as $\hat{\lambda}=\frac{\sum_{i\neq j}Var(\hat{r}_{ij})}{\sum_{i\neq j}\hat{r}_{ij}^{2}}$, where $\hat{r}_{ij}$ is the $ij^{th}$ element of $\hat{R}_{1}$ which is the 1-step-ahead sample correlation matrix.

\subsection{Base forecasting approaches}
\label{sec:BFA}

In hierarchical forecasting, forecasts are generated for each series at every node of the hierarchy individually using different forecasting modelling approaches. These
individual forecasts are referred to as base forecasts. In this study we consider Naive method, Seasonal naive method (SNAIVE), average method, AutoRegressive Integrated Moving Average (ARIMA) approach, and Exponential Smoothing method (ETS) to generate base forecasts.

In Naive approach, forecasts are equal to the last observed value in the series. $$\hat{y}_{T+h|T}=y_{T}$$ SNAIVE approach considers that the forecasts are equal to the last observed value from the same season of the year. That means, forecasts for time $T+h$ is $$\hat{y}_{T+h|T}=y_{T+h-m(k+1)}$$ where $m$ is the seasonal period, $k$ is the integer part of $\frac{(h-1)}{m}$. In the average method, the forecasts of all future values are equal to the average (or mean) of the available data. If $y_{1},y_{2},\dots,y_{T}$ denotes the available data, then  $$\hat{y}_{T+h|T}=\bar{y}=(y_{1}+y_{2}+\dots+y_{T})/{T}$$ Auto-correlations in the data are described by ARIMA models. ARIMA model is a collection of auto-regressive and moving average components which captures the effect of past observations and error terms. ARIMA model comprises with 3 terms: $p, d, q$ where $p$ is the order of AR term, $q$ is the order of MA term, and $d$ is the number of differences required to make the time series into stationary. AR term considers the linear regression model which depends only on lagged term as predictors $$y_{t}=c+\phi_{1}y_{t-1}+\phi_{2}y_{t-2}+\dots+\phi_{p}y_{t-p}+\epsilon_{t}$$ MA term in ARIMA model describes the effect of lagged error terms $$y_{t}=c+\epsilon_{t}+\theta_{1}\epsilon_{t-1}+\theta_{2}\epsilon_{t-2}+\dots+\theta_{q}\epsilon_{t-q}$$ Exponential Smoothing models are constructed by considering the weighted averages of past observations in which high weights are assigned to the most recent observation. It was proposed by \citet{holt2004forecasting} and \citet{winters1960forecasting}. Statistical frame work of exponential smoothing methods was proposed by \citet{hyndman2002state} are illustrated by Table \ref{table:figurechapter38}.
\begin{longtable}[t]{lccc}
\caption{\label{table:figurechapter38}Exponential Smoothing methods}\\
\toprule
\multicolumn{1}{c}{ } & \multicolumn{3}{c}{Seasonal Component} \\
\cmidrule(l{3pt}r{3pt}){2-4}
Trend Component & N(None) & A(Additive) & M(Multiplicative)\\
\midrule
\endfirsthead
\caption[]{Exponential Smoothing methods \textit{(continued)}}\\
\toprule
\multicolumn{1}{c}{ } & \multicolumn{3}{c}{Seasonal Component} \\
\cmidrule(l{3pt}r{3pt}){2-4}
Trend Component & N(None) & A(Additive) & M(Multiplicative)\\
\midrule
\endhead

\endfoot
\bottomrule
\endlastfoot
N (None) & (N,N) & (N,A) & (N,M)\\
A (Additive) & (A,N) & (A,A) & (A,M)\\
Ad (Additive damped) & (Ad,N) & (Ad,A) & (Ad,M)\\
Multiplicative (M) & (M,N) & (M,A) & (M,M)\\
Md (Multiplicative damped) & (Md,N) & (Md,A) & (Md,M)\\*
\end{longtable}
For example, cell (N, N) corresponds to simple exponential smoothing method. Two possible state space models (additive errors or multiplicative errors) are available for each of the methods in Table \ref{table:figurechapter38}.

\subsection{Forecast performance evaluation}
\label{sec:FPE}
We evaluate the performance of base forecasting approaches against hierarchical forecasting approaches. Many accuracy measures have been proposed to check the forecast accuracy \citep{hyndman2006another}. However, many of these proposed accuracy measures are not applicable due to the reasons such as scale dependency, producing undefined or infinite value and giving misleading results. Therefore, in this study we use Mean Absolute Scale Error (MASE) to obtain the accuracy of forecasting methods. For $h$-step-ahead forecasts, $$MAE^{a}=\frac{1}{h}\Sigma_{j=1}^{h}|y_{j}-\hat{y}_{j}|,$$ is the mean absolute error of the forecast method $a$, $y_{j}$ and $\hat{y}_{j}$ are the actual and forecast values at period $j$ respectively. Then, $$MASE=\frac{MAE^{a}}{Q},$$ where $Q=\frac{1}{T-m}\Sigma_{t=1}^{T}|y_{t}-y_{t-m}|$ is the scaling factor where $m$ is the sampling frequency per year \citep{athanasopoulos2017forecasting}.

\subsection{Data and modelling strategies}
\label{sec:DCSL}
\begin{wrapfigure}{r}{0.4\textwidth}
    \centering
    \includegraphics[width=0.4\textwidth]{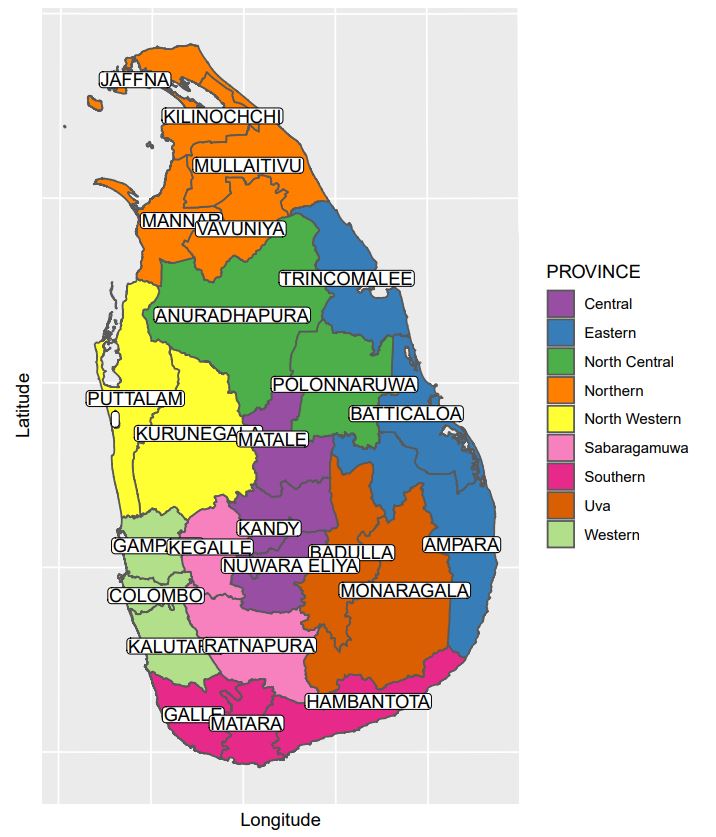}
    \caption{Sri Lanka map}
    \label{fig:figurechapter22}
\end{wrapfigure}

\begin{figure}[h!]
\centering
  \includegraphics[scale=0.8]{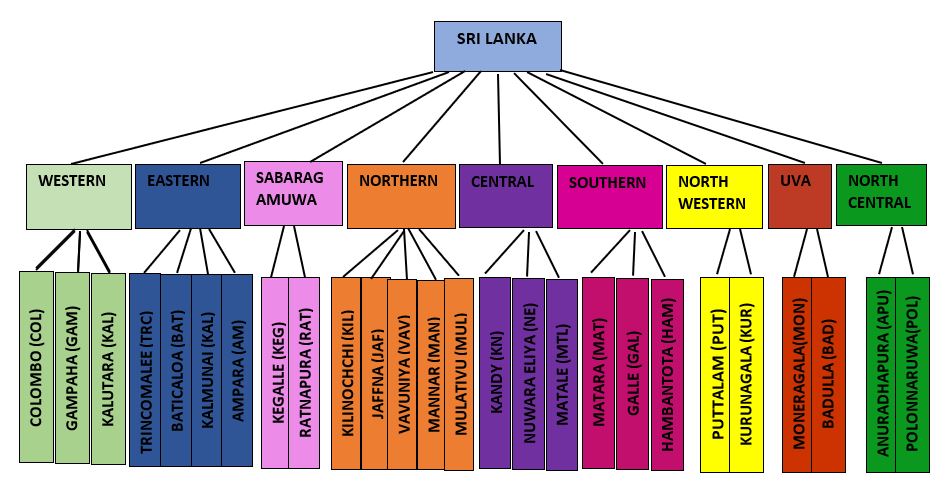}
  \caption{ Spatial hierarchical structure}
  \label{fig:figurechapter23}
\end{figure}

\begin{figure}[h!]
\centering
  \includegraphics[scale=0.6]{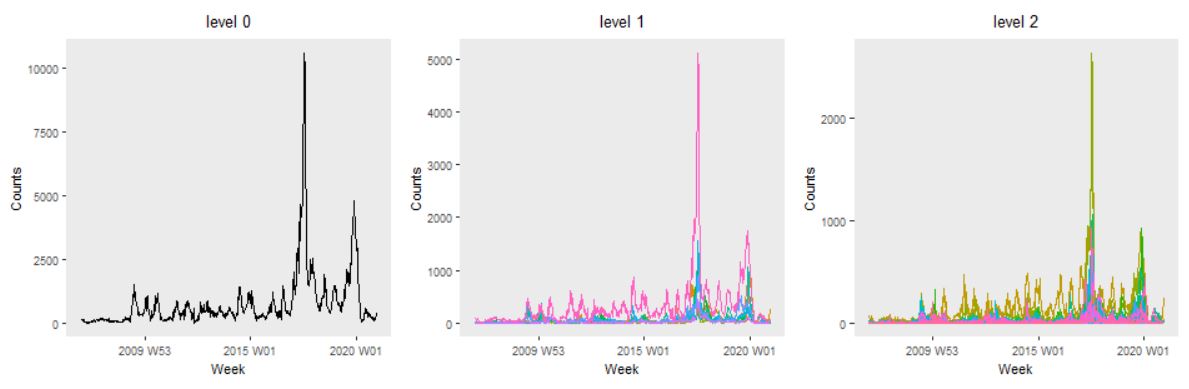}
  \caption{ Time series of Dengue counts in Sri Lanka for different aggregated levels of the spatial hierarchical structure}
  \label{fig:figurechapter24}
\end{figure}

Weekly notified dengue cases in 26 districts of Sri Lanka were obtained from Epidemiological Unit, Ministry of Health, Sri Lanka for the period of $52^{nd}$ week of 2006 to $52^{nd}$ week of 2020. This study considers the three levels of geographic divisions in Sri Lanka. In the first level, Sri Lanka is divided into 9 provinces such as Central (CN), Eastern (ET), North Central (NC), Nothern (NT), North Western (NW), Sabaragamuwa (SB), Southern (ST), UVA, and Western (WT) provinces. In the second level, these 9 provinces are further classified into 26 districts. In addition, level 0 is the most aggregate level or dengue counts of the whole country. The geographic structure is illustrated as a map in Figure \ref{fig:figurechapter22} and the corresponding spatial hierarchical structure is illustrated by Figure \ref{fig:figurechapter23}.\\
As shown in Figure \ref{fig:figurechapter24}, the series of each level of the spatial hierarchical structure shown in Figure \ref{fig:figurechapter23} reveal different time series patterns, respectively.
According to Figure \ref{fig:figurechapter24}, time series are different in scales and time series features. In Figure \ref{fig:figurechapter24}, the panel corresponds to level 0 demonstrates dengue counts of Sri Lanka and it indicates two significant spikes during each year due to the monsoon season of Sri Lanka. A notable upward trend pattern is also visible. When we focus on level 1 and level 2 of hierarchy, these pattern variations are less prominent. Bottom level series are much noisier than other levels. Hence, it is more challenging to model the series of bottom level. Generating coherent accurate forecasts for the series at each levels are extremely important to successfully plan the curative and preventive programs.\par
\begin{figure}[h!]
\centering
  \includegraphics[scale=0.4]{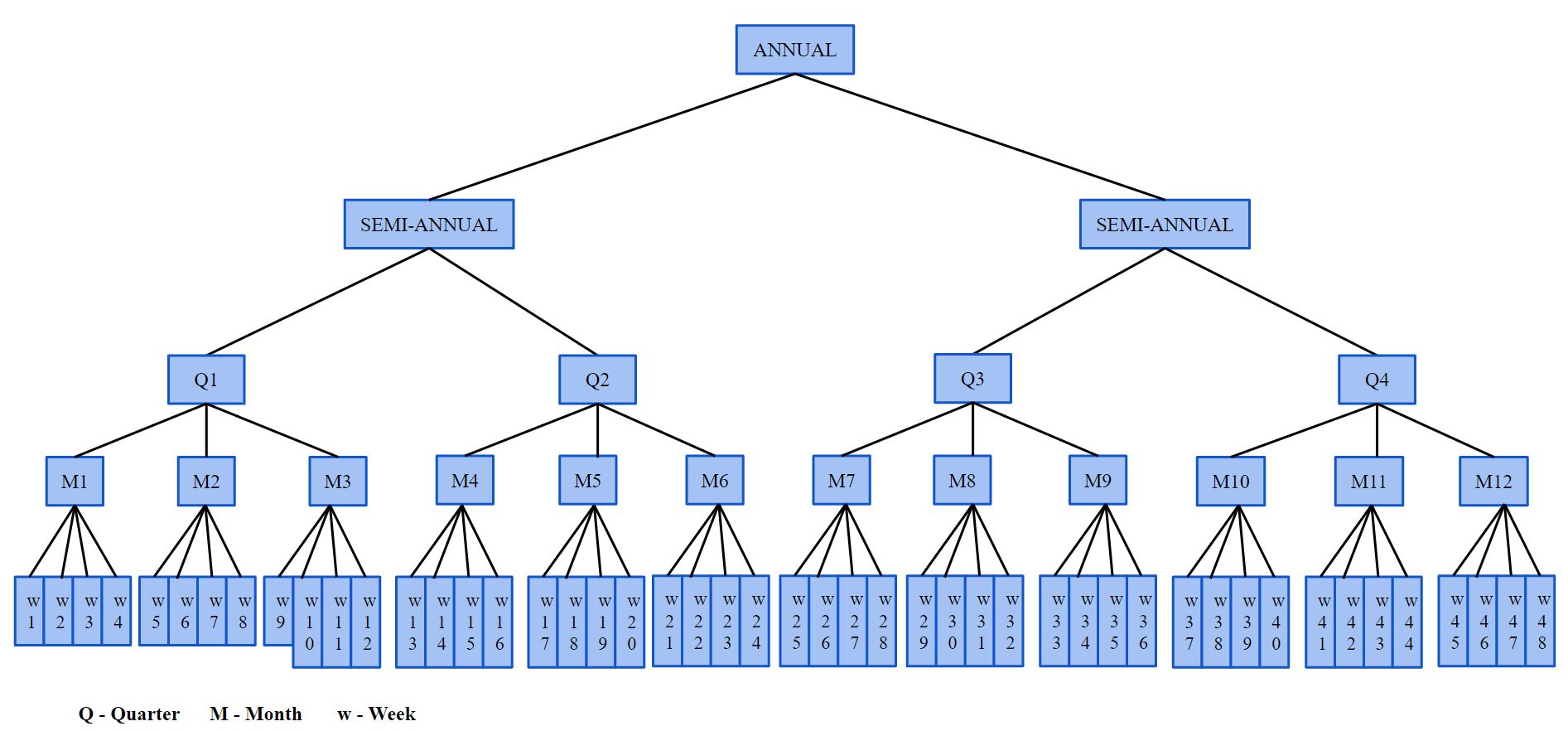}
  \caption{Temporal hierarchical structure}
  \label{fig:figurechapter25}
\end{figure}
\begin{figure}[h!]
\centering
  \includegraphics[scale=0.4]{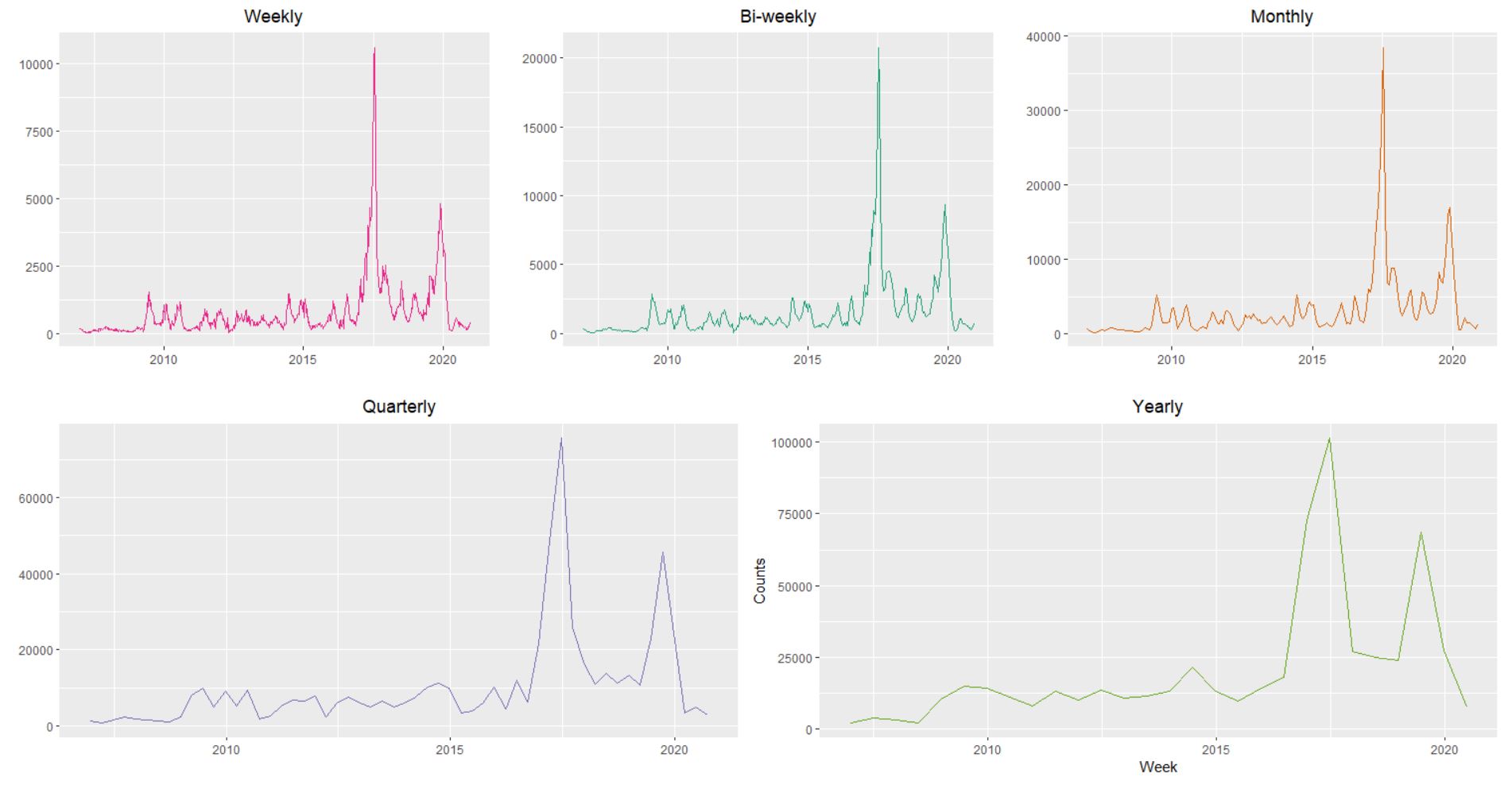}
  \caption{Time series of Dengue counts in Sri Lanka according to temporal hierarchical structure}
  \label{fig:figurechapter26}
\end{figure}

In addition, dengue curative and preventive programs require different strategies for different time period. Hence, this study focuses on the temporal structure of data as demonstrated by Figure \ref{fig:figurechapter25}.
Figure \ref{fig:figurechapter26} represents the time series dynamic of dengue incidence for different aggregated levels in Sri Lanka. According to Figure \ref{fig:figurechapter26}, the weekly time series depicts the highest seasonal variation. Then, bi-weekly series have been omitted the negligible seasonality variations. Next, monthly series have demonstrated the important seasonality component and a small trend component. Finally, the yearly time series depicts upward trend of the series.

Before fitting the model, we separate the data set into training set and test sets. This study considers four different training sets such as 2006-2019, 2016-2019, 2006-2018, 2016-2018 and test sets are 2020, 2019 respectively. Furthermore, weekly, monthly, quarterly, and yearly frequencies are taken into the account in all these training and test sets. All statistical analysis related to spatial hierarchical forecasting was performed by fabletools package \citep{hyndman2018forecasting} whereas temporal hierarchical forecasting was implemented by thief package in R \citep{hyndman2018thief}.

In the case of average, NAIVE, SNAIVE approaches, adjusting base forecasts are not necessary to obtain reconciled forecasts due to the fact that forecasts already follow the aggregate constraints associated with spatial hierarchical structure. In temporal hierarchical structure, forecasts of average and SNAIVE approach only adhere to constraints associated with temporal generalities.  Hence, forecast reconciliation is necessary for ARIMA, ETS, and NAIVE methods in temporal hierarchical forecasting whereas forecast reconciliation is needed for ARIMA and ETS method in spatial hierarchical forecasting.

\section{Results}

\subsection{Forecasting results}
\label{sec:res1}

\begin{figure}[h!]
\centering
  \includegraphics[scale=0.75]{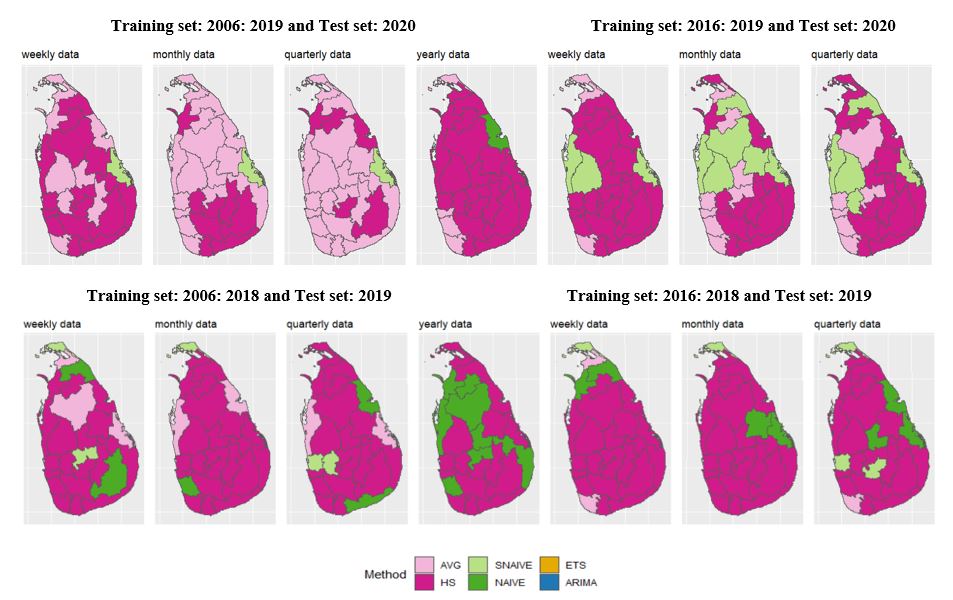}
  \caption{The best forecasting method of level 2 according to training set}
  \label{fig:figurechapter27}
\end{figure}

\begin{figure}[h!]
\centering
  \includegraphics[scale=0.75]{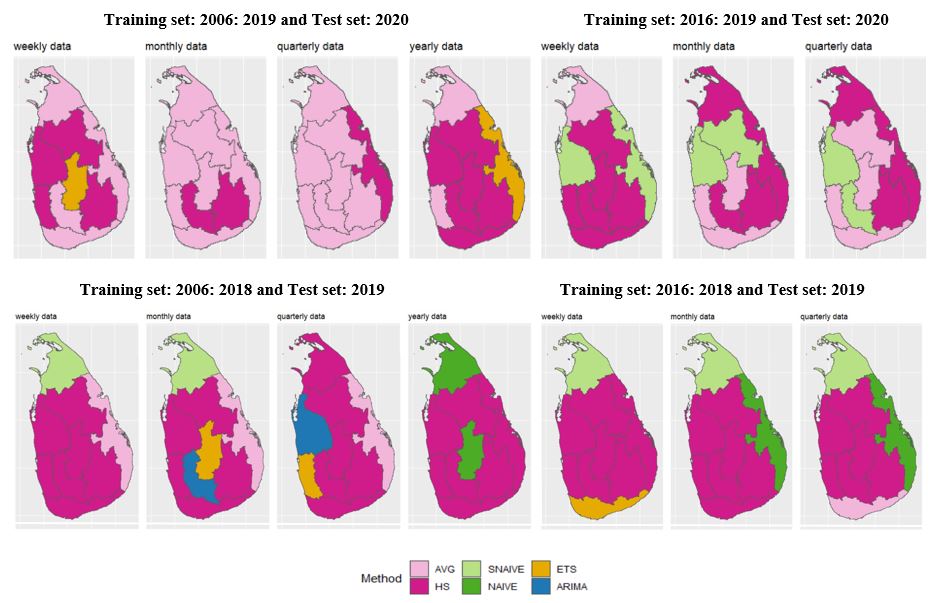}
  \caption{The best forecasting method of level 1 according to training set}
  \label{fig:figurechapter28}
\end{figure}

\begin{figure}[H]
\centering
  \includegraphics[width=0.5\textwidth]{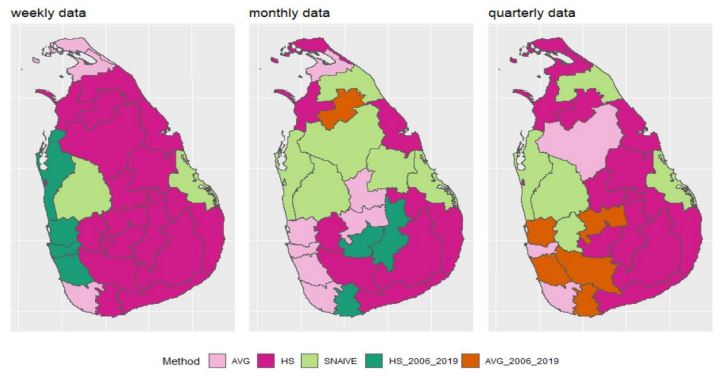}
  \caption{District with the best forecasting method for 2020}
  \label{fig:figurechapter212n}
\end{figure}

\begin{figure}[H]
\centering
  \includegraphics[width=0.5\textwidth]{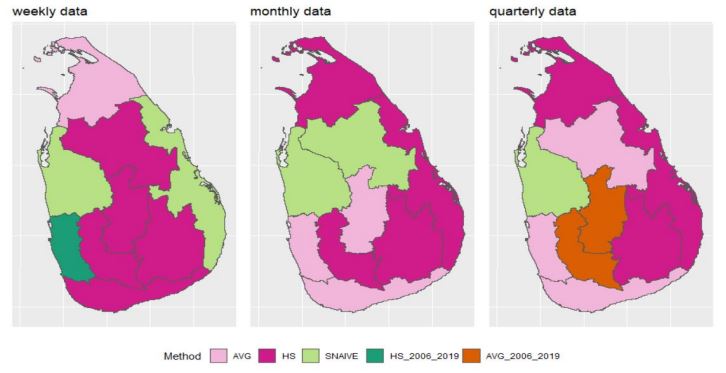}
  \caption{Provinces with the best forecasting method for 2020}
  \label{fig:figurechapter212nn}
\end{figure}

We compare the forecasts accuracy for all series in the spatial and temporal hierarchical structure. We compare the accuracy of forecasts generated based of hierarchical forecasting approach against Average method (AVG), NAIVE, SNAIVE, ETS and ARIMA. The computed MASE values over the test sets of all series in the spatial hierarchical structure are summarized from Table \ref{tab:tnew} to  Table \ref{tab:foo2}. We then identified the best forecasting method for all series at each level of the of the hierarchy. The corresponding results are shown in  Figure \ref{fig:figurechapter27} - Figure \ref{fig:figurechapter28}.  According to the results it can be seen that when generating forecasts for 2019, for most of the geographical regions hierarchical forecasts give the best forecast. However, when generating forecasts for the year 2020, the average method gives the best forecasts for Colombo district, Kalutara district (level 2) and Western province (level 1). This could be due to ongoing COVID-19 pandemic. Sometimes healthcare providers faced challenges to distinguish COVID-19 from dengue. Hence, there were some unusual pattern in dengue cases in 2020 specially for population density area.
For the country-level series, the lowest MASE was given by the forecasts generated from average method based on the training sets 2006-2019 (TS1) and 2016-2019 (TS2)  whereas the best forecasting approach for the training sets 2006-2018 (TS3) and 2016-2018 (TS4) is hierarchical time series forecasting (HS) approach. Similarly, we generated weekly, monthly and quarterly forecasts based on temporal hierarchical structure and compared the accuracy of results with other benchmark approaches. The results are are summarized in Table \ref{tab:misc}. Furthermore, Figure \ref{fig:figurechapter212n} and \ref{fig:figurechapter212nn} show the best forecasting method for weekly, monthly and quarterly for districts and provinces respectively. For most of the geographical regions forecasts generated based on the hierarchical approach gives the best forecasts.


\begin{table}[!h]
\caption{\label{tab:misc}The best forecasting approaches for temporal granualities according to training set}
\centering
\fontsize{9}{11}\selectfont
\begin{tabular}[t]{lcccccccc}
\toprule
\multicolumn{1}{c}{ } & \multicolumn{8}{c}{Training set} \\
\cmidrule(l{3pt}r{3pt}){2-9}
\multicolumn{1}{c}{ } & \multicolumn{2}{c}{2006-2018} & \multicolumn{2}{c}{2016-2018} & \multicolumn{2}{c}{2006-2019} & \multicolumn{2}{c}{2016-2019} \\
\cmidrule(l{3pt}r{3pt}){2-3} \cmidrule(l{3pt}r{3pt}){4-5} \cmidrule(l{3pt}r{3pt}){6-7} \cmidrule(l{3pt}r{3pt}){8-9}
Series & Method & MASE & Method & MASE & Method & MASE & Method & MASE\\
\midrule
\cellcolor{gray!6}{Annual} & \cellcolor{gray!6}{HS} & \cellcolor{gray!6}{0.26} & \cellcolor{gray!6}{HS} & \cellcolor{gray!6}{0.02} & \cellcolor{gray!6}{AVG} & \cellcolor{gray!6}{0.20} & \cellcolor{gray!6}{HS} & \cellcolor{gray!6}{0.09}\\
Semi-annual & HS & 1.28 & HS & 0.29 & AVG & 0.61 & HS & 0.14\\
\cellcolor{gray!6}{Quarterly} & \cellcolor{gray!6}{HS} & \cellcolor{gray!6}{1.17} & \cellcolor{gray!6}{HS} & \cellcolor{gray!6}{0.29} & \cellcolor{gray!6}{AVG} & \cellcolor{gray!6}{0.97} & \cellcolor{gray!6}{HS} & \cellcolor{gray!6}{0.21}\\
Monthly & HS & 1.17 & HS & 0.31 & AVG & 0.96 & HS & 0.23\\
\cellcolor{gray!6}{Bi-Weekly} & \cellcolor{gray!6}{HS} & \cellcolor{gray!6}{1.18} & \cellcolor{gray!6}{HS} & \cellcolor{gray!6}{0.31} & \cellcolor{gray!6}{AVG} & \cellcolor{gray!6}{0.97} & \cellcolor{gray!6}{HS} & \cellcolor{gray!6}{0.24}\\
\addlinespace
Weekly & HS & 1.17 & HS & 0.32 & AVG & 0.96 & HS & 0.24\\
\bottomrule
\end{tabular}
\end{table}

Based on the above results the best forecasting approaches in each series of the spatial hierarchical structure and temporal hierarchical structure are not limited to single approach. According to No-Free-Lunch theorem, there is no single method that fits all situations  \citep{kang2017visualising}. Hence, this study attempts to identify the reasons for the best forecasting approach variations based on time series features.

\subsection{Visualization of model performance}
\label{sec:VMP}

\begin{figure}[H]
\centering
  \includegraphics[width=0.95\textwidth]{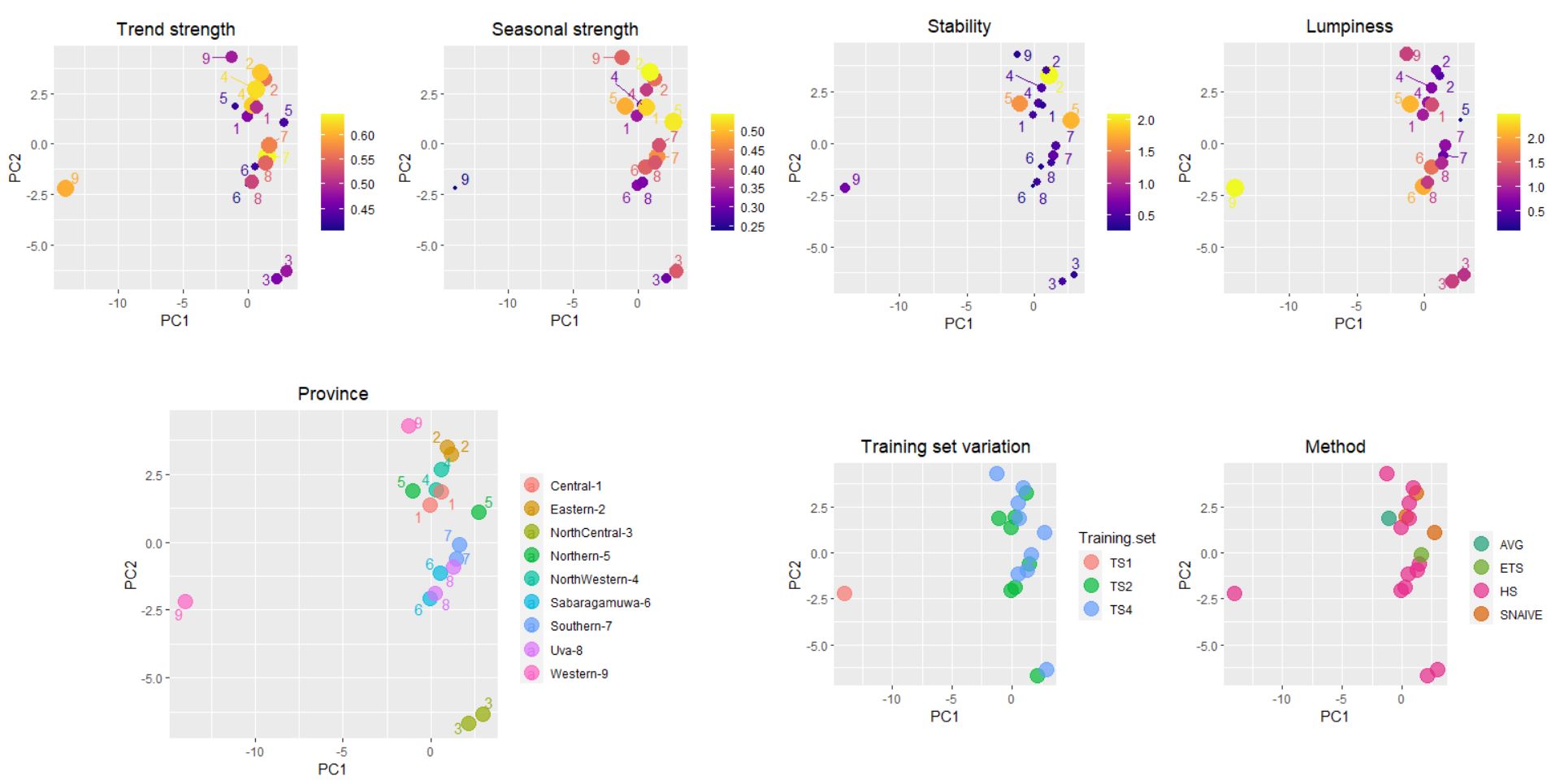}
  \caption{Model performance visualization for weekly province-wise series of spatial hierarchical structure}
  \label{fig:figurechapter212}
\end{figure}

\begin{figure}[H]
\centering
  \includegraphics[width=0.95\textwidth]{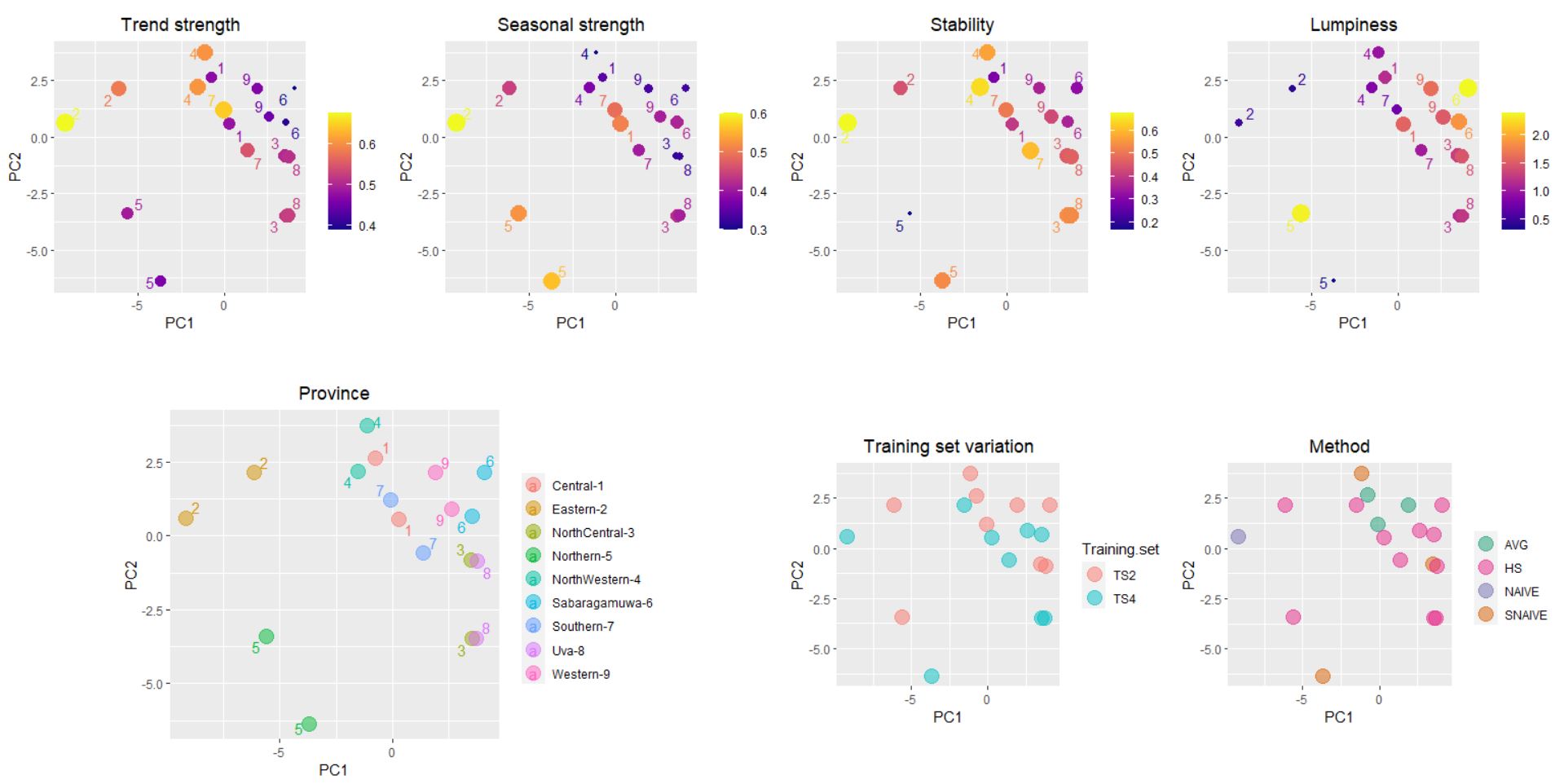}
  \caption{Model performance visualization for monthly province-wise series of spatial hierarchical structure}
  \label{fig:figurechapter213}
\end{figure}

\begin{figure}[H]
\centering
  \includegraphics[width=0.95\textwidth]{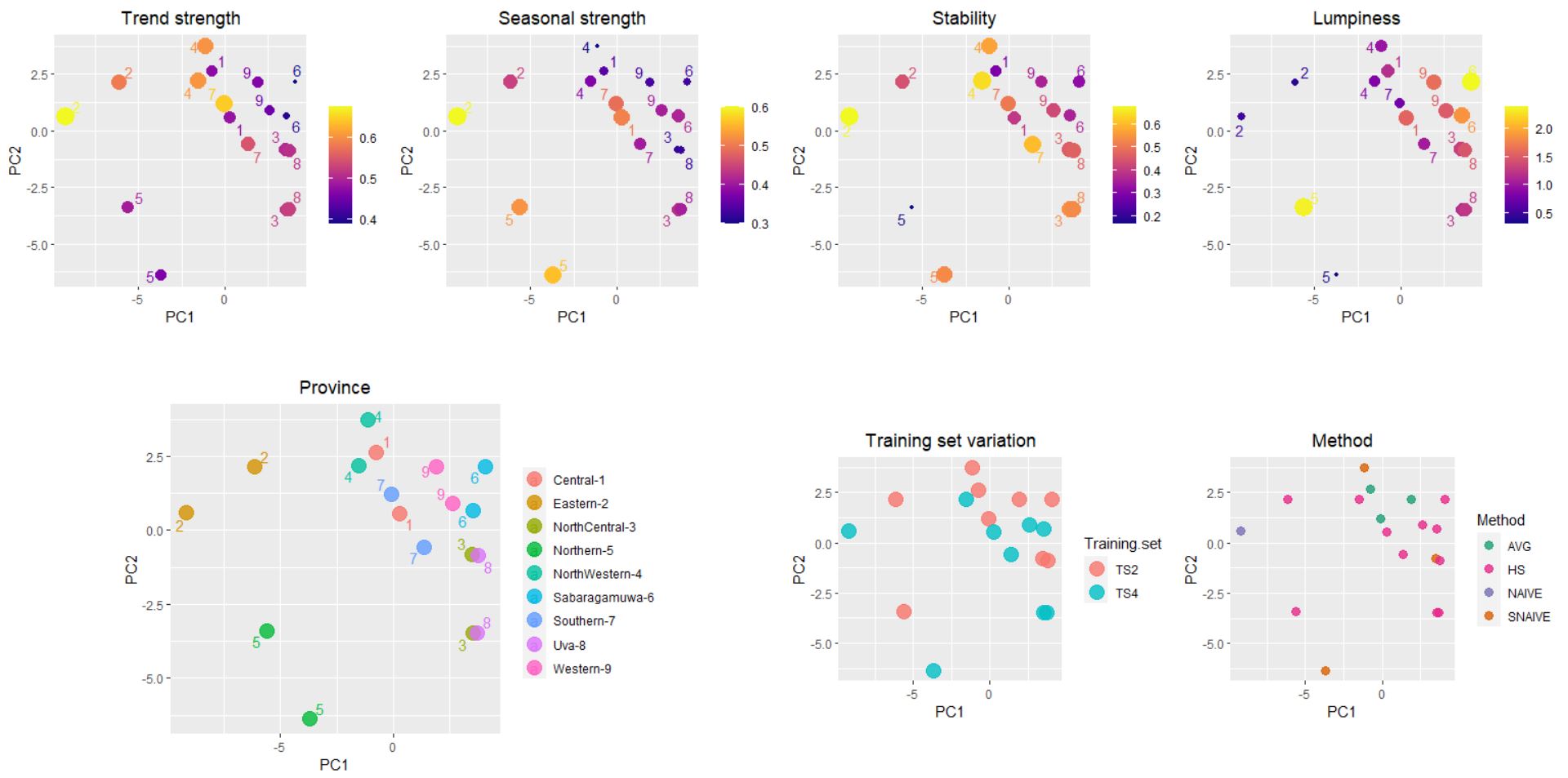}
  \caption{Model performance visualization for quarterly province-wise series of spatial hierarchical structure}
  \label{fig:figurechapter214}
\end{figure}

\begin{figure}[H]
\centering
  \includegraphics[width=0.95\textwidth]{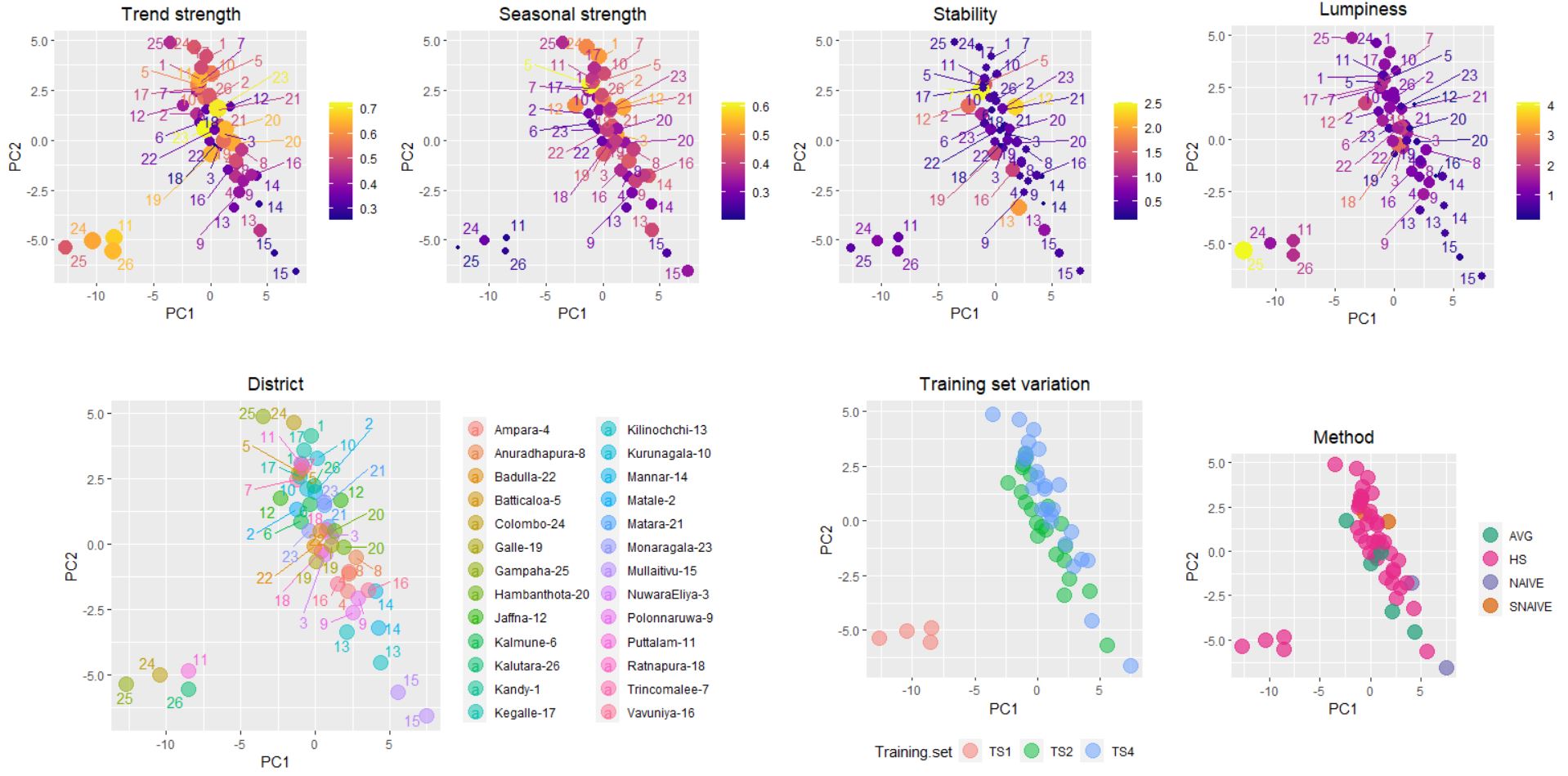}
  \caption{Model performance visualization for weekly district-wise series of spatial hierarchical structure}
  \label{fig:figurechapter215}
\end{figure}

\begin{figure}[H]
\centering
  \includegraphics[width=0.95\textwidth]{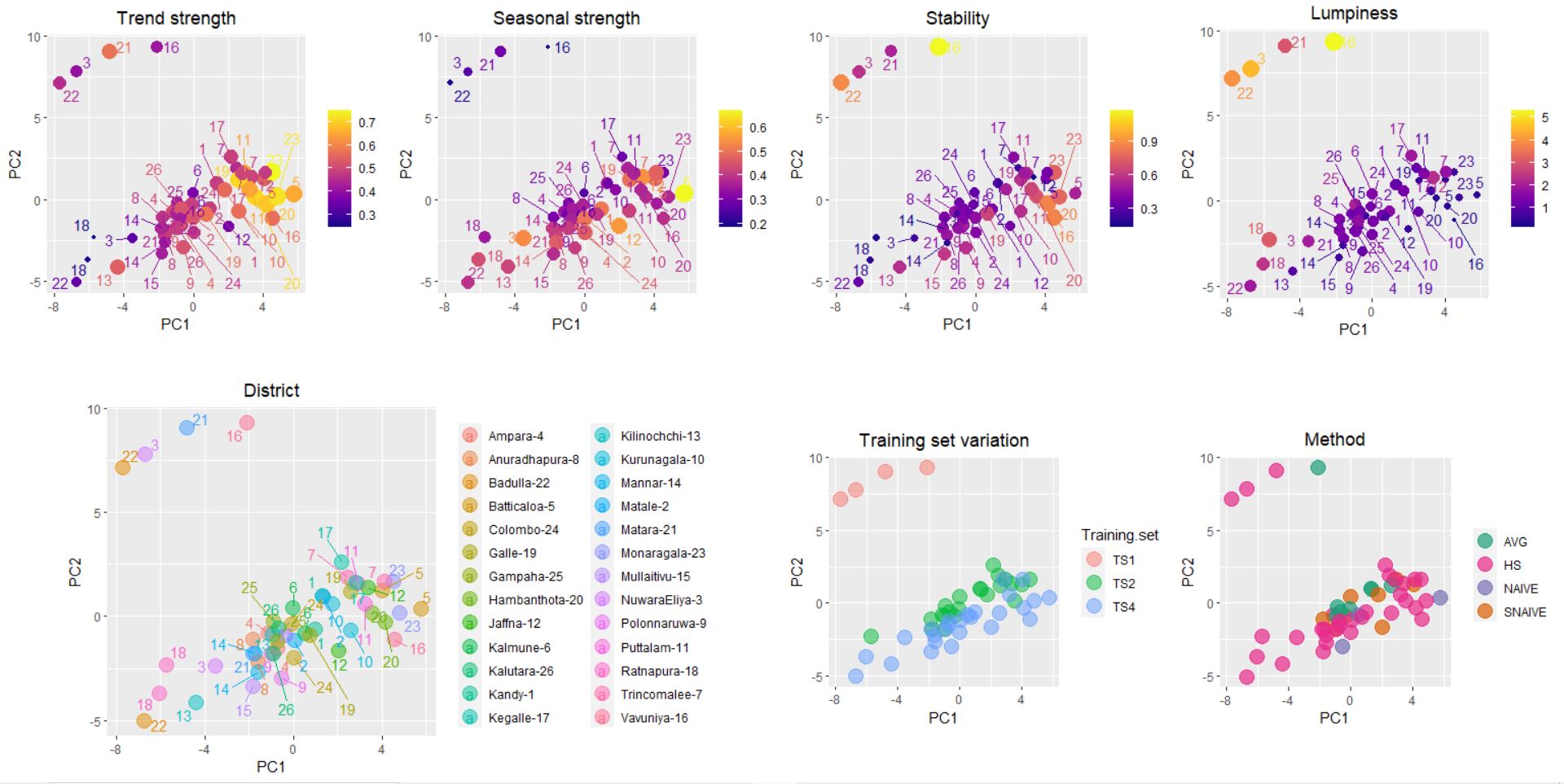}
  \caption{Model performance visualization for monthly district-wise series of spatial hierarchical structure}
  \label{fig:figurechapter216}
\end{figure}

\begin{figure}[H]
\centering
  \includegraphics[width=0.95\textwidth]{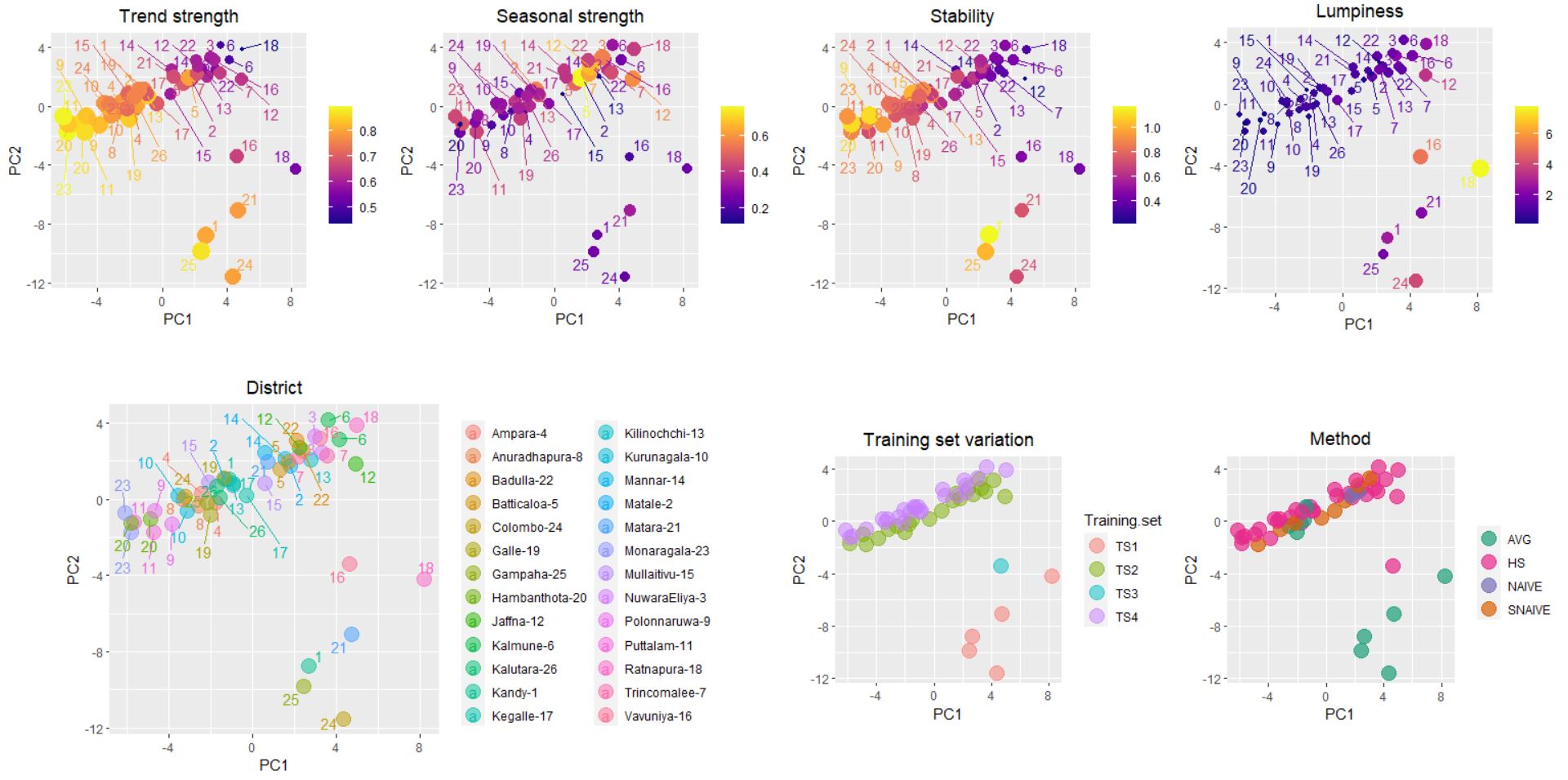}
  \caption{Model performance visualization for quarterly district-wise series of spatial hierarchical structure}
  \label{fig:figurechapter217}
\end{figure}

In Section \ref{sec:res1}, we identified the best forecasting approaches corresponding to weekly, monthly, quarterly, and yearly series for the years 2019 and 2020 under four different training sets based on the spatial or temporal hierarchical structure. Hence, this section attempts to identify the reasons for the forecasting model performance variations based on feature-based visualization. In order to obtain feature-based visualization, first features of time series are extracted from the training set corresponding to the best forecasting method training set. The features are computed based on the feasts package in R \citep{feasts}. We computed 48 features from each series. They are strength of seasonality, strength of trend, number of flat spots, statistic based on Lagrange Multiplier test, optimal $\lambda$ value of Box-cox transformation, seasonal peak year, seasonal trough year, spikiness, linearity, curvature, first and tenth autocorrelation coefficient of the remainder component, first and tenth autocorrelation coefficient from the original data, longest flat spot length, average interval between non-zero observations, squared coefficient of variation of non-zero observations, proportion of data which starts with zero and ends with zero, autocorrelation function at lag 1 and lag 10 from the first differenced data, autocorrelation function at lag 1 and lag 10 from the second differenced data, autocorrelation function at lag 1 from first seasonal lag data, hurst coefficient, spectral entropy, Box-Pierce statistic, p-value from Box-Pierce statistic, Ljung-Box statistic, p-value from Ljung-Box statistic, partial autocorrelation function at lag 5, partial autocorrelation function at lag 5 from first differenced data, partial autocorrelation function at lag 5 from second differenced data, partial autocorrelation at the first seasonal lag, KPSS statistic, p-value of KPSS test, Phillips-Perron statistic, p-value from Phillips-Perron test, required number of differences, required number of seasonal differences, stability, lumpiness, the largest mean shift between two consecutive sliding windows, index at which the largest mean shift occurs, the largest variance shift between two consecutive sliding windows, index at which the largest variance shift occurs, the largest distributional shift, index at which the largest distributional shift occurs, and number of times time series crosses the median. A detailed description of the features is provided in  \citet{talagala2018meta}. Next, the principal component analysis (PCA) was performed to reduce the dimension of features. Furthermore,  PCA was conducted separately for district-wise or province-wise weekly, monthly, and quarterly series. Principal component 1 versus principal component 2 is plotted to visualize feature variation according to the modeling approach and the corresponding visualizations are presented from Figures \ref{fig:figurechapter212} to \ref{fig:figurechapter217}. The points in the resulting scatter plots are coloured according to four features: i) strength of trend, ii) strength of seasonality, iii) stability and iv) lumpiness. The two features stability and lumpiness are calculated based on non-overlapping windows of the time series. Then stability is measures based on the variance of the means, while lumpiness is measured based on the variance of the variances. The reason for selecting these features are they are very easy to interpret. In addition to that the points are coloured according to their province/ district name, training set corresponds to best forecasting method and best forecasting method (forecasting method that gives the lowest MASE).

According to Figure \ref{fig:figurechapter212}, only for Nothern province the best forecasts were given by SNAIVE and average (AVG) method based on the training sets TS4 and TS2 respectively. According to the feature visualization we can see the strength of trend corresponds to the Nothern province is very low compared to other series. Furthermore, the stability corresponds to the Nothern province is considerably higher. According to Figure \ref{fig:figurechapter213}, Eastern province and Sabaragamuwa province are located far away from the other provinces due to differences in the features. The best forecasting method for Eastern province is NAIVE approach. This could be due to differences in trend, seasonality and lumpiness compared to other series.  In Figure \ref{fig:figurechapter212} to \ref{fig:figurechapter217}, for all series that generates best forecasts with NAIVE or SNAIVE have low value for lumpiness, high value for stability, strength of trend and strength of seasonality. In addition to that in we see outlying stay points that are isolated from other points tends to give best forecasts based on NAIVE, SNAIVE or Average method. 



.
\section{Conclusion}
\label{sec:CON}

In this study, we have applied hierarchical time series forecasting approaches to forecast dengue incidence in Sri Lanka. Hierarchical time series forecasting approaches are extremely important to obtain optimal decisions which nicely match with the decisions across the levels of hierarchical structure. In hierarchical forecasting, first, we need to obtain forecasts of each series without considering the hierarchical structure. This study obtain the forecasts by ARIMA, ETS, NAIVE, SNAIVE, and average approaches which are referred to as base forecasts. Then, these forecasts are adjusted in order to preserve the coherency of forecasts. However, hierarchical time series forecasting approaches do not provide the lowest MASE values in all situations of spatial hierarchical series.  Furthermore, the number of observations in the training set does not impact the accuracy improvement. According to the feature-based visualization of spatial hierarchical structure, SNAIVE and NAIVE approaches are not the most accurate method if a series contains a high lumpiness value. Furthermore, the hierarchical time series forecasting approach indicates the best performance in all ranges of seasonality, trend, stability, and lumpiness feature values. In the case of the high lumpiness value, the average method also provides accurate performance. According to the results of temporal hierarchical forecasting, the hierarchical time series forecasting approach is the best approach for the years 2019 and 2020. According to feature-based visualization we observe that the choice of the best forecasting method depends of the features of the time series. We leave the study of cross-temporal hierarchical forecasting as a future research. We have developed a shiny app to disseminate the results of the analysis with the general public. The app is available at \url{https://github.com/SamudraMadushani/DengueSriLankaApp}. All data are shared through an open source R package called DISC (DISease Counts) which is available at \url{https://github.com/SMART-Research/DISC}. The maps are generated using ceylon R package available at \url{https://github.com/thiyangt/ceylon}.

\bigskip

\begin{center}
{\large\bf APPENDIX}
\end{center}
We present the results corresponds to base and reconciled forecasts for four different training and test sets in spatial hierarchical forecasting from Table \ref{tab:tnew} to  \ref{tab:foo}. Base forecasting approaches such as ARIMA, ETS, NAIVE, SNAIVE, and Average are denoted by arm, snv, nve, ets, avg respectively. Hierarchical time series forecasting approaches such as bottom up, top down, ordinary least square estimator using variance scaling, structural scaling, one-step covariance matrix when base forecasting approach is ARIMA are denoted by bup, top, ols, mit, var, stc, and cov.

\begin{landscape}\begin{table}[!ht]
\caption{\label{tab:tnew}MASE from training set: 2006-2019 for weekly data}
\centering
\fontsize{8}{8}\selectfont
\begin{tabular}[t]{r>{}r>{}r>{}r>{}r>{}r>{}r>{}rrrrr>{}r>{}r>{}rr>{}rr>{}r>{}r}
\toprule
\multicolumn{1}{c}{ } & \multicolumn{5}{c}{Base Forecasting} & \multicolumn{7}{c}{HTS based on ARIMA} & \multicolumn{7}{c}{HTS based on ETS} \\
\cmidrule(l{3pt}r{3pt}){2-6} \cmidrule(l{3pt}r{3pt}){7-13} \cmidrule(l{3pt}r{3pt}){14-20}
series & arm & snv & nve & ets & avg & bup & top & ols & mit & var & stc & cov & ebup & etop & eols & emit & evar & estc & ecov\\
\midrule
\cellcolor{gray!6}{SL} & \cellcolor{gray!6}{3.97} & \cellcolor{gray!6}{2.14} & \cellcolor{gray!6}{4.07} & \cellcolor{gray!6}{4.49} & \cellcolor{gray!6}{\textbf{0.90}} & \cellcolor{gray!6}{3.96} & \cellcolor{gray!6}{3.97} & \cellcolor{gray!6}{3.95} & \cellcolor{gray!6}{3.85} & \cellcolor{gray!6}{3.92} & \cellcolor{gray!6}{3.93} & \cellcolor{gray!6}{3.10} & \cellcolor{gray!6}{3.90} & \cellcolor{gray!6}{4.49} & \cellcolor{gray!6}{4.36} & \cellcolor{gray!6}{3.51} & \cellcolor{gray!6}{3.75} & \cellcolor{gray!6}{3.91} & \cellcolor{gray!6}{1.31}\\
\addlinespace[0.3em]
\multicolumn{20}{l}{\textbf{Province}}\\
\hspace{1em}CN & 2.16 & 1.97 & 3.79 & \textbf{0.54} & 0.62 & 3.47 & 2.24 & 2.55 & 2.87 & 2.91 & 2.83 & 2.35 & 1.87 & 0.76 & 1.58 & 1.28 & 1.28 & 1.40 & 1.67\\
\hspace{1em}\cellcolor{gray!6}{ET} & \cellcolor{gray!6}{7.70} & \cellcolor{gray!6}{2.18} & \cellcolor{gray!6}{7.70} & \cellcolor{gray!6}{7.70} & \cellcolor{gray!6}{\textbf{1.93}} & \cellcolor{gray!6}{7.68} & \cellcolor{gray!6}{8.01} & \cellcolor{gray!6}{7.84} & \cellcolor{gray!6}{7.70} & \cellcolor{gray!6}{7.70} & \cellcolor{gray!6}{7.75} & \cellcolor{gray!6}{7.40} & \cellcolor{gray!6}{8.53} & \cellcolor{gray!6}{10.19} & \cellcolor{gray!6}{8.97} & \cellcolor{gray!6}{8.23} & \cellcolor{gray!6}{8.19} & \cellcolor{gray!6}{8.57} & \cellcolor{gray!6}{8.97}\\
\hspace{1em}NC & 2.83 & 1.28 & 2.87 & 2.84 & 0.84 & 2.65 & 2.91 & 3.10 & 2.53 & 2.72 & 2.80 & \textbf{0.48} & 2.84 & 3.81 & 6.12 & 2.44 & 2.87 & 3.65 & 1.56\\
\bottomrule
\end{tabular}
\end{table}
\end{landscape}

\newpage

\begin{landscape}\begin{table}[!ht]

\caption{\label{tab:foo7}MASE from training set: 2006-2019 for monthly data}
\centering
\fontsize{8}{8}\selectfont
\begin{tabular}[t]{lr>{}r>{}r>{}r>{}rrr>{}rrr>{}rr>{}r>{}rrrr>{}rr}
\toprule
\multicolumn{1}{c}{ } & \multicolumn{5}{c}{Base Forecasting} & \multicolumn{7}{c}{HTS based on ARIMA} & \multicolumn{7}{c}{HTS based on ETS} \\
\cmidrule(l{3pt}r{3pt}){2-6} \cmidrule(l{3pt}r{3pt}){7-13} \cmidrule(l{3pt}r{3pt}){14-20}
series & arm & snv & nve & ets & avg & bup & top & ols & mit & var & stc & cov & ebup & etop & eols & emit & evar & estc & ecov\\
\midrule
\cellcolor{gray!6}{SL} & \cellcolor{gray!6}{2.25} & \cellcolor{gray!6}{2.18} & \cellcolor{gray!6}{4.35} & \cellcolor{gray!6}{4.11} & \cellcolor{gray!6}{\textbf{0.85}} & \cellcolor{gray!6}{2.50} & \cellcolor{gray!6}{2.25} & \cellcolor{gray!6}{2.40} & \cellcolor{gray!6}{4.51} & \cellcolor{gray!6}{2.94} & \cellcolor{gray!6}{2.79} & \cellcolor{gray!6}{4.14} & \cellcolor{gray!6}{4.53} & \cellcolor{gray!6}{4.11} & \cellcolor{gray!6}{4.24} & \cellcolor{gray!6}{5.73} & \cellcolor{gray!6}{4.81} & \cellcolor{gray!6}{4.63} & \cellcolor{gray!6}{3.66}\\
\addlinespace[0.3em]
\multicolumn{20}{l}{\textbf{Province}}\\
\hspace{1em}CN & 4.92 & 2.05 & 4.59 & 5.22 & \textbf{0.60} & 4.80 & 3.06 & 3.94 & 5.72 & 4.81 & 4.60 & 3.30 & 4.64 & 4.06 & 4.28 & 5.97 & 4.85 & 4.69 & 3.58\\
\hspace{1em}\cellcolor{gray!6}{ET} & \cellcolor{gray!6}{7.28} & \cellcolor{gray!6}{2.17} & \cellcolor{gray!6}{6.39} & \cellcolor{gray!6}{13.83} & \cellcolor{gray!6}{\textbf{1.66}} & \cellcolor{gray!6}{1.96} & \cellcolor{gray!6}{4.62} & \cellcolor{gray!6}{4.96} & \cellcolor{gray!6}{4.87} & \cellcolor{gray!6}{4.42} & \cellcolor{gray!6}{4.20} & \cellcolor{gray!6}{8.12} & \cellcolor{gray!6}{6.33} & \cellcolor{gray!6}{10.53} & \cellcolor{gray!6}{11.11} & \cellcolor{gray!6}{10.28} & \cellcolor{gray!6}{10.12} & \cellcolor{gray!6}{9.59} & \cellcolor{gray!6}{11.17}\\
\hspace{1em}NC & 2.67 & 1.36 & 3.15 & 3.26 & \textbf{0.85} & 1.74 & 1.64 & 1.56 & 3.70 & 2.08 & 1.44 & 5.91 & 2.53 & 2.55 & 1.47 & 4.40 & 2.87 & 2.15 & 5.07\\
\bottomrule
\end{tabular}
\end{table}
\end{landscape}

\newpage

\begin{landscape}\begin{table}[!ht]

\caption{\label{tab:foo8}MASE from training set: 2006-2019 for quarterly data}
\centering
\fontsize{8}{8}\selectfont
\begin{tabular}[t]{lr>{}r>{}r>{}r>{}rrr>{}r>{}r>{}r>{}r>{}rrrrrr}
\toprule
\multicolumn{1}{c}{ } & \multicolumn{5}{c}{Base Forecasting} & \multicolumn{6}{c}{HTS based on ARIMA} & \multicolumn{6}{c}{HTS based on ETS} \\
\cmidrule(l{3pt}r{3pt}){2-6} \cmidrule(l{3pt}r{3pt}){7-12} \cmidrule(l{3pt}r{3pt}){13-18}
series & arm & snv & nve & ets & avg & bup & top & ols & mit & var & stc & ebup & etop & eols & emit & evar & estc\\
\midrule
\cellcolor{gray!6}{SL} & \cellcolor{gray!6}{2.65} & \cellcolor{gray!6}{2.17} & \cellcolor{gray!6}{4.38} & \cellcolor{gray!6}{4.64} & \cellcolor{gray!6}{\textbf{0.69}} & \cellcolor{gray!6}{2.82} & \cellcolor{gray!6}{2.65} & \cellcolor{gray!6}{2.68} & \cellcolor{gray!6}{2.87} & \cellcolor{gray!6}{2.85} & \cellcolor{gray!6}{2.81} & \cellcolor{gray!6}{3.76} & \cellcolor{gray!6}{4.64} & \cellcolor{gray!6}{4.57} & \cellcolor{gray!6}{3.63} & \cellcolor{gray!6}{3.95} & \cellcolor{gray!6}{4.18}\\
\addlinespace[0.3em]
\multicolumn{18}{l}{\textbf{Province}}\\
\hspace{1em}CN & 1.84 & 1.94 & 5.41 & 5.55 & \textbf{0.31} & 1.99 & 1.72 & 1.67 & 1.99 & 1.92 & 1.84 & 5.07 & 6.19 & 5.89 & 4.81 & 5.31 & 5.53\\
\hspace{1em}\cellcolor{gray!6}{ET} & \cellcolor{gray!6}{1.80} & \cellcolor{gray!6}{2.19} & \cellcolor{gray!6}{3.44} & \cellcolor{gray!6}{2.30} & \cellcolor{gray!6}{2.03} & \cellcolor{gray!6}{1.82} & \cellcolor{gray!6}{1.45} & \cellcolor{gray!6}{\textbf{1.43}} & \cellcolor{gray!6}{1.80} & \cellcolor{gray!6}{1.78} & \cellcolor{gray!6}{1.53} & \cellcolor{gray!6}{2.08} & \cellcolor{gray!6}{2.42} & \cellcolor{gray!6}{2.53} & \cellcolor{gray!6}{2.23} & \cellcolor{gray!6}{2.16} & \cellcolor{gray!6}{2.33}\\
\hspace{1em}NC & 2.68 & 1.44 & 2.68 & 2.83 & \textbf{0.81} & 2.43 & 2.36 & 1.67 & 2.52 & 2.53 & 2.31 & 2.62 & 3.17 & 4.74 & 2.34 & 2.71 & 3.41\\
\bottomrule
\end{tabular}
\end{table}
\end{landscape}

\newpage

\begin{landscape}\begin{table}[!ht]

\caption{\label{tab:foo811}MASE from training set: 2006-2019 for yearly data}
\centering
\fontsize{8}{8}\selectfont
\begin{tabular}[t]{l>{}r>{}r>{}r>{}r>{}r>{}r>{}r>{}r>{}r>{}r>{}rr>{}r>{}rr>{}r}
\toprule
\multicolumn{1}{c}{ } & \multicolumn{4}{c}{Base Forecasting} & \multicolumn{6}{c}{HTS based on ARIMA} & \multicolumn{6}{c}{HTS based on ETS} \\
\cmidrule(l{3pt}r{3pt}){2-5} \cmidrule(l{3pt}r{3pt}){6-11} \cmidrule(l{3pt}r{3pt}){12-17}
series & arm & nve & ets & avg & bup & top & ols & mit & var & stc & ebup & etop & eols & emit & evar & estc\\
\midrule
\cellcolor{gray!6}{SL} & \cellcolor{gray!6}{\textbf{0.24}} & \cellcolor{gray!6}{2.16} & \cellcolor{gray!6}{1.29} & \cellcolor{gray!6}{\textbf{0.24}} & \cellcolor{gray!6}{0.75} & \cellcolor{gray!6}{\textbf{0.24}} & \cellcolor{gray!6}{0.34} & \cellcolor{gray!6}{0.84} & \cellcolor{gray!6}{0.80} & \cellcolor{gray!6}{0.68} & \cellcolor{gray!6}{1.50} & \cellcolor{gray!6}{1.29} & \cellcolor{gray!6}{1.32} & \cellcolor{gray!6}{1.54} & \cellcolor{gray!6}{1.49} & \cellcolor{gray!6}{1.43}\\
\addlinespace[0.3em]
\multicolumn{17}{l}{\textbf{Province}}\\
\hspace{1em}CN & 0.10 & 1.91 & 0.92 & 0.10 & 0.10 & 0.52 & 0.68 & \textbf{0.02} & 0.12 & 0.30 & 0.87 & 0.74 & 0.72 & 0.89 & 0.88 & 0.83\\
\hspace{1em}\cellcolor{gray!6}{ET} & \cellcolor{gray!6}{1.68} & \cellcolor{gray!6}{0.52} & \cellcolor{gray!6}{\textbf{0.18}} & \cellcolor{gray!6}{2.12} & \cellcolor{gray!6}{1.97} & \cellcolor{gray!6}{2.37} & \cellcolor{gray!6}{2.79} & \cellcolor{gray!6}{1.84} & \cellcolor{gray!6}{1.92} & \cellcolor{gray!6}{2.28} & \cellcolor{gray!6}{0.28} & \cellcolor{gray!6}{0.47} & \cellcolor{gray!6}{0.55} & \cellcolor{gray!6}{0.19} & \cellcolor{gray!6}{0.26} & \cellcolor{gray!6}{0.38}\\
\hspace{1em}NC & 0.33 & 1.14 & 0.72 & 0.33 & 0.33 & 0.12 & 2.23 & 0.40 & 0.32 & 0.33 & 1.40 & 0.58 & \textbf{0.11} & 1.23 & 1.20 & 0.84\\
\bottomrule
\end{tabular}
\end{table}
\end{landscape}

\newpage

\begin{landscape}\begin{table}[!ht]

\caption{\label{tab:foo3}MASE from training set: 2016-2019 for weekly data}
\centering
\fontsize{8}{8}\selectfont
\begin{tabular}[t]{l>{}r>{}r>{}r>{}r>{}r>{}r>{}r>{}rrr>{}r>{}r>{}r>{}rrrrr>{}r}
\toprule
\multicolumn{1}{c}{ } & \multicolumn{5}{c}{Base Forecasting} & \multicolumn{7}{c}{HTS based on ARIMA} & \multicolumn{7}{c}{HTS based on ETS} \\
\cmidrule(l{3pt}r{3pt}){2-6} \cmidrule(l{3pt}r{3pt}){7-13} \cmidrule(l{3pt}r{3pt}){14-20}
series & arm & snv & nve & ets & avg & bup & top & ols & mit & var & stc & cov & ebup & etop & eols & emit & evar & estc & ecov\\
\midrule
\cellcolor{gray!6}{SL} & \cellcolor{gray!6}{0.66} & \cellcolor{gray!6}{0.76} & \cellcolor{gray!6}{1.45} & \cellcolor{gray!6}{1.86} & \cellcolor{gray!6}{\textbf{0.64}} & \cellcolor{gray!6}{0.85} & \cellcolor{gray!6}{0.66} & \cellcolor{gray!6}{0.68} & \cellcolor{gray!6}{0.89} & \cellcolor{gray!6}{0.83} & \cellcolor{gray!6}{0.79} & \cellcolor{gray!6}{1.36} & \cellcolor{gray!6}{1.68} & \cellcolor{gray!6}{1.86} & \cellcolor{gray!6}{1.85} & \cellcolor{gray!6}{1.81} & \cellcolor{gray!6}{1.73} & \cellcolor{gray!6}{1.77} & \cellcolor{gray!6}{1.83}\\
\addlinespace[0.3em]
\multicolumn{20}{l}{\textbf{Province}}\\
\hspace{1em}CN & 0.45 & 0.77 & 1.49 & 2.12 & 0.50 & 0.45 & 0.32 & \textbf{0.29} & 0.50 & 0.44 & 0.38 & 1.04 & 1.90 & 2.24 & 2.17 & 2.08 & 2.00 & 2.06 & 2.34\\
\hspace{1em}\cellcolor{gray!6}{ET} & \cellcolor{gray!6}{1.33} & \cellcolor{gray!6}{\textbf{0.75}} & \cellcolor{gray!6}{2.64} & \cellcolor{gray!6}{3.29} & \cellcolor{gray!6}{0.85} & \cellcolor{gray!6}{0.86} & \cellcolor{gray!6}{1.05} & \cellcolor{gray!6}{1.03} & \cellcolor{gray!6}{1.04} & \cellcolor{gray!6}{1.04} & \cellcolor{gray!6}{0.99} & \cellcolor{gray!6}{1.01} & \cellcolor{gray!6}{2.94} & \cellcolor{gray!6}{3.48} & \cellcolor{gray!6}{3.35} & \cellcolor{gray!6}{3.07} & \cellcolor{gray!6}{3.09} & \cellcolor{gray!6}{3.19} & \cellcolor{gray!6}{2.76}\\
\hspace{1em}NC & 0.58 & 0.53 & 1.19 & 1.17 & 0.58 & 0.59 & 0.46 & \textbf{0.19} & 0.68 & 0.58 & 0.39 & 1.90 & 1.17 & 1.24 & 1.63 & 1.23 & 1.18 & 1.32 & 2.03\\
\bottomrule
\end{tabular}
\end{table}
\end{landscape}

\newpage

\begin{landscape}\begin{table}[!ht]

\caption{\label{tab:foo4}MASE from training set: 2016-2019 for monthly data}
\centering
\fontsize{8}{8}\selectfont
\begin{tabular}[t]{l>{}r>{}r>{}r>{}r>{}r>{}r>{}rr>{}r>{}rr>{}r>{}rrrrrr>{}r}
\toprule
\multicolumn{1}{c}{ } & \multicolumn{5}{c}{Base Forecasting} & \multicolumn{7}{c}{HTS based on ARIMA} & \multicolumn{7}{c}{HTS based on ETS} \\
\cmidrule(l{3pt}r{3pt}){2-6} \cmidrule(l{3pt}r{3pt}){7-13} \cmidrule(l{3pt}r{3pt}){14-20}
series & arm & snv & nve & ets & avg & bup & top & ols & mit & var & stc & cov & ebup & etop & eols & emit & evar & estc & ecov\\
\midrule
\cellcolor{gray!6}{SL} & \cellcolor{gray!6}{0.70} & \cellcolor{gray!6}{0.77} & \cellcolor{gray!6}{1.54} & \cellcolor{gray!6}{1.41} & \cellcolor{gray!6}{\textbf{0.63}} & \cellcolor{gray!6}{\textbf{0.63}} & \cellcolor{gray!6}{0.70} & \cellcolor{gray!6}{0.70} & \cellcolor{gray!6}{0.73} & \cellcolor{gray!6}{0.66} & \cellcolor{gray!6}{0.67} & \cellcolor{gray!6}{1.13} & \cellcolor{gray!6}{1.79} & \cellcolor{gray!6}{1.41} & \cellcolor{gray!6}{1.43} & \cellcolor{gray!6}{1.56} & \cellcolor{gray!6}{1.59} & \cellcolor{gray!6}{1.57} & \cellcolor{gray!6}{1.86}\\
\addlinespace[0.3em]
\multicolumn{20}{l}{\textbf{Province}}\\
\hspace{1em}CN & 0.69 & 0.80 & 1.80 & 2.28 & \textbf{0.50} & 0.72 & 0.70 & 0.72 & 0.78 & 0.71 & 0.72 & 1.22 & 2.20 & 2.12 & 2.12 & 2.10 & 2.20 & 2.15 & 2.15\\
\hspace{1em}\cellcolor{gray!6}{ET} & \cellcolor{gray!6}{0.71} & \cellcolor{gray!6}{0.74} & \cellcolor{gray!6}{2.17} & \cellcolor{gray!6}{1.48} & \cellcolor{gray!6}{0.75} & \cellcolor{gray!6}{\textbf{0.54}} & \cellcolor{gray!6}{0.73} & \cellcolor{gray!6}{0.70} & \cellcolor{gray!6}{0.62} & \cellcolor{gray!6}{0.61} & \cellcolor{gray!6}{0.64} & \cellcolor{gray!6}{1.28} & \cellcolor{gray!6}{2.02} & \cellcolor{gray!6}{1.42} & \cellcolor{gray!6}{1.43} & \cellcolor{gray!6}{1.71} & \cellcolor{gray!6}{1.73} & \cellcolor{gray!6}{1.62} & \cellcolor{gray!6}{1.41}\\
\hspace{1em}NC & 0.58 & \textbf{0.55} & 1.27 & 1.62 & 0.58 & 0.59 & 0.60 & 0.71 & 0.66 & 0.59 & 0.64 & 1.08 & 1.63 & 1.50 & 1.14 & 1.48 & 1.62 & 1.36 & 1.76\\
\bottomrule
\end{tabular}
\end{table}
\end{landscape}

\newpage

\begin{landscape}\begin{table}[!ht]

\caption{\label{tab:foo5}MASE for training set: 2016-2019 for quarterly data}
\centering
\fontsize{8}{8}\selectfont
\begin{tabular}[t]{lrrrrrrrrrrrrrrrrr}
\toprule
\multicolumn{1}{c}{ } & \multicolumn{5}{c}{Base Forecasting} & \multicolumn{6}{c}{HTS based on ARIMA} & \multicolumn{6}{c}{HTS based on ETS} \\
\cmidrule(l{3pt}r{3pt}){2-6} \cmidrule(l{3pt}r{3pt}){7-12} \cmidrule(l{3pt}r{3pt}){13-18}
series & arm & snv & nve & ets & avg & bup & top & ols & mit & var & stc & ebup & etop & eols & emit & evar & estc\\
\midrule
\cellcolor{gray!6}{SL} & \cellcolor{gray!6}{0.65} & \cellcolor{gray!6}{0.77} & \cellcolor{gray!6}{1.55} & \cellcolor{gray!6}{1.92} & \cellcolor{gray!6}{0.52} & \cellcolor{gray!6}{0.60} & \cellcolor{gray!6}{0.65} & \cellcolor{gray!6}{0.65} & \cellcolor{gray!6}{0.68} & \cellcolor{gray!6}{0.62} & \cellcolor{gray!6}{0.63} & \cellcolor{gray!6}{1.44} & \cellcolor{gray!6}{1.92} & \cellcolor{gray!6}{1.87} & \cellcolor{gray!6}{1.49} & \cellcolor{gray!6}{1.54} & \cellcolor{gray!6}{1.66}\\
\addlinespace[0.3em]
\multicolumn{18}{l}{\textbf{Province}}\\
\hspace{1em}CN & 0.68 & 0.77 & 2.14 & 2.58 & 0.43 & 0.59 & 0.70 & 0.68 & 0.71 & 0.63 & 0.64 & 2.18 & 3.03 & 2.79 & 2.30 & 2.36 & 2.52\\
\hspace{1em}\cellcolor{gray!6}{ET} & \cellcolor{gray!6}{0.57} & \cellcolor{gray!6}{0.76} & \cellcolor{gray!6}{1.19} & \cellcolor{gray!6}{0.73} & \cellcolor{gray!6}{0.71} & \cellcolor{gray!6}{0.70} & \cellcolor{gray!6}{0.53} & \cellcolor{gray!6}{0.54} & \cellcolor{gray!6}{0.63} & \cellcolor{gray!6}{0.65} & \cellcolor{gray!6}{0.61} & \cellcolor{gray!6}{0.76} & \cellcolor{gray!6}{0.80} & \cellcolor{gray!6}{0.97} & \cellcolor{gray!6}{0.81} & \cellcolor{gray!6}{0.76} & \cellcolor{gray!6}{0.87}\\
\hspace{1em}NC & 0.58 & 0.59 & 1.09 & 1.50 & 0.52 & 0.76 & 0.60 & 0.74 & 0.73 & 0.70 & 0.71 & 1.48 & 1.77 & 2.87 & 1.40 & 1.49 & 1.93\\
\bottomrule
\end{tabular}
\end{table}
\end{landscape}

\newpage

\begin{landscape}\begin{table}[!ht]

\caption{\label{tab:tneww}MASE from training set: 2006-2018 for weekly data}
\centering
\fontsize{8}{8}\selectfont
\begin{tabular}[t]{lrrrrrrrrrrrrrrrrrrr}
\toprule
\multicolumn{1}{c}{ } & \multicolumn{5}{c}{Base Forecasting} & \multicolumn{7}{c}{HTS based on ARIMA} & \multicolumn{7}{c}{HTS based on ETS} \\
\cmidrule(l{3pt}r{3pt}){2-6} \cmidrule(l{3pt}r{3pt}){7-13} \cmidrule(l{3pt}r{3pt}){14-20}
series & arm & snv & nve & ets & avg & bup & top & ols & mit & var & stc & cov & ebup & etop & eols & emit & evar & estc & ecov\\
\midrule
\cellcolor{gray!6}{SL} & \cellcolor{gray!6}{1.29} & \cellcolor{gray!6}{1.44} & \cellcolor{gray!6}{1.34} & \cellcolor{gray!6}{1.19} & \cellcolor{gray!6}{1.62} & \cellcolor{gray!6}{1.34} & \cellcolor{gray!6}{1.29} & \cellcolor{gray!6}{1.30} & \cellcolor{gray!6}{1.33} & \cellcolor{gray!6}{1.34} & \cellcolor{gray!6}{1.33} & \cellcolor{gray!6}{1.27} & \cellcolor{gray!6}{1.35} & \cellcolor{gray!6}{1.19} & \cellcolor{gray!6}{1.21} & \cellcolor{gray!6}{1.36} & \cellcolor{gray!6}{1.34} & \cellcolor{gray!6}{1.30} & \cellcolor{gray!6}{1.84}\\
\addlinespace[0.3em]
\multicolumn{20}{l}{\textbf{Province}}\\
\hspace{1em}CN & 2.44 & 1.82 & 1.90 & 2.59 & 1.83 & 1.97 & 2.42 & 2.18 & 2.11 & 2.15 & 2.15 & 1.82 & 2.30 & 2.59 & 2.11 & 2.41 & 2.40 & 2.29 & 2.73\\
\hspace{1em}\cellcolor{gray!6}{ET} & \cellcolor{gray!6}{1.43} & \cellcolor{gray!6}{2.36} & \cellcolor{gray!6}{1.39} & \cellcolor{gray!6}{1.41} & \cellcolor{gray!6}{1.33} & \cellcolor{gray!6}{1.36} & \cellcolor{gray!6}{1.53} & \cellcolor{gray!6}{1.51} & \cellcolor{gray!6}{1.38} & \cellcolor{gray!6}{1.39} & \cellcolor{gray!6}{1.42} & \cellcolor{gray!6}{1.35} & \cellcolor{gray!6}{1.42} & \cellcolor{gray!6}{1.62} & \cellcolor{gray!6}{1.67} & \cellcolor{gray!6}{1.43} & \cellcolor{gray!6}{1.42} & \cellcolor{gray!6}{1.52} & \cellcolor{gray!6}{1.48}\\
\hspace{1em}NC & 0.76 & 1.03 & 0.78 & 0.77 & 0.77 & 0.75 & 0.75 & 0.81 & 0.75 & 0.75 & 0.74 & 0.77 & 0.77 & 0.80 & 1.38 & 0.78 & 0.76 & 0.81 & 2.41\\
\bottomrule
\end{tabular}
\end{table}
\end{landscape}

\newpage

\begin{landscape}\begin{table}[!ht]

\caption{\label{tab:foo71}MASE from training set: 2006-2018for monthly data}
\centering
\fontsize{8}{8}\selectfont
\begin{tabular}[t]{lrrrrrrrrrrrrrrrrrrr}
\toprule
\multicolumn{1}{c}{ } & \multicolumn{5}{c}{Base Forecasting} & \multicolumn{7}{c}{HTS based on ARIMA} & \multicolumn{7}{c}{HTS based on ETS} \\
\cmidrule(l{3pt}r{3pt}){2-6} \cmidrule(l{3pt}r{3pt}){7-13} \cmidrule(l{3pt}r{3pt}){14-20}
series & arm & snv & nve & ets & avg & bup & top & ols & mit & var & stc & cov & ebup & etop & eols & emit & evar & estc & ecov\\
\midrule
\cellcolor{gray!6}{SL} & \cellcolor{gray!6}{1.34} & \cellcolor{gray!6}{1.45} & \cellcolor{gray!6}{1.35} & \cellcolor{gray!6}{1.29} & \cellcolor{gray!6}{1.60} & \cellcolor{gray!6}{1.39} & \cellcolor{gray!6}{1.34} & \cellcolor{gray!6}{1.34} & \cellcolor{gray!6}{1.33} & \cellcolor{gray!6}{1.36} & \cellcolor{gray!6}{1.35} & \cellcolor{gray!6}{1.66} & \cellcolor{gray!6}{1.30} & \cellcolor{gray!6}{1.29} & \cellcolor{gray!6}{1.28} & \cellcolor{gray!6}{1.31} & \cellcolor{gray!6}{1.29} & \cellcolor{gray!6}{1.28} & \cellcolor{gray!6}{1.80}\\
\addlinespace[0.3em]
\multicolumn{20}{l}{\textbf{Province}}\\
\hspace{1em}CN & 1.87 & 1.83 & 1.82 & 1.68 & 1.88 & 1.87 & 1.88 & 1.87 & 1.91 & 1.87 & 1.87 & 2.45 & 1.80 & 1.71 & 1.74 & 1.76 & 1.72 & 1.73 & 2.19\\
\hspace{1em}\cellcolor{gray!6}{ET} & \cellcolor{gray!6}{1.60} & \cellcolor{gray!6}{2.48} & \cellcolor{gray!6}{1.55} & \cellcolor{gray!6}{1.55} & \cellcolor{gray!6}{1.31} & \cellcolor{gray!6}{2.02} & \cellcolor{gray!6}{1.55} & \cellcolor{gray!6}{1.65} & \cellcolor{gray!6}{1.85} & \cellcolor{gray!6}{1.82} & \cellcolor{gray!6}{1.81} & \cellcolor{gray!6}{2.35} & \cellcolor{gray!6}{1.53} & \cellcolor{gray!6}{1.34} & \cellcolor{gray!6}{1.33} & \cellcolor{gray!6}{1.51} & \cellcolor{gray!6}{1.54} & \cellcolor{gray!6}{1.44} & \cellcolor{gray!6}{1.32}\\
\hspace{1em}NC & 0.84 & 1.03 & 0.72 & 0.76 & 0.76 & 0.71 & 0.85 & 0.91 & 0.74 & 0.74 & 0.76 & 0.85 & 0.74 & 0.58 & 0.51 & 0.61 & 0.64 & 0.53 & 1.34\\
\bottomrule
\end{tabular}
\end{table}
\end{landscape}

\newpage

\begin{landscape}\begin{table}[!ht]

\caption{\label{tab:foo82}MASE from training set: 2006-2018 for quarterly data}
\centering
\fontsize{8}{8}\selectfont
\begin{tabular}[t]{lrrrrrrrrrrrrrrrrr}
\toprule
\multicolumn{1}{c}{ } & \multicolumn{5}{c}{Base Forecasting} & \multicolumn{6}{c}{HTS based on ARIMA} & \multicolumn{6}{c}{HTS based on ETS} \\
\cmidrule(l{3pt}r{3pt}){2-6} \cmidrule(l{3pt}r{3pt}){7-12} \cmidrule(l{3pt}r{3pt}){13-18}
series & arm & snv & nve & ets & avg & bup & top & ols & mit & var & stc & ebup & etop & eols & emit & evar & estc\\
\midrule
\cellcolor{gray!6}{SL} & \cellcolor{gray!6}{1.46} & \cellcolor{gray!6}{1.52} & \cellcolor{gray!6}{1.46} & \cellcolor{gray!6}{1.43} & \cellcolor{gray!6}{1.73} & \cellcolor{gray!6}{1.46} & \cellcolor{gray!6}{1.46} & \cellcolor{gray!6}{1.46} & \cellcolor{gray!6}{1.41} & \cellcolor{gray!6}{1.45} & \cellcolor{gray!6}{1.46} & \cellcolor{gray!6}{1.55} & \cellcolor{gray!6}{1.43} & \cellcolor{gray!6}{1.43} & \cellcolor{gray!6}{1.41} & \cellcolor{gray!6}{1.50} & \cellcolor{gray!6}{1.46}\\
\addlinespace[0.3em]
\multicolumn{18}{l}{\textbf{Province}}\\
\hspace{1em}CN & 1.93 & 1.90 & 1.93 & 1.85 & 1.93 & 2.26 & 1.94 & 1.96 & 2.06 & 2.07 & 2.02 & 1.85 & 1.87 & 1.85 & 1.80 & 1.85 & 1.84\\
\hspace{1em}\cellcolor{gray!6}{ET} & \cellcolor{gray!6}{1.97} & \cellcolor{gray!6}{2.72} & \cellcolor{gray!6}{1.42} & \cellcolor{gray!6}{1.72} & \cellcolor{gray!6}{1.42} & \cellcolor{gray!6}{1.88} & \cellcolor{gray!6}{1.92} & \cellcolor{gray!6}{1.93} & \cellcolor{gray!6}{1.94} & \cellcolor{gray!6}{1.91} & \cellcolor{gray!6}{1.93} & \cellcolor{gray!6}{2.40} & \cellcolor{gray!6}{1.69} & \cellcolor{gray!6}{1.75} & \cellcolor{gray!6}{2.24} & \cellcolor{gray!6}{2.18} & \cellcolor{gray!6}{1.96}\\
\hspace{1em}NC & 0.79 & 0.90 & 0.80 & 0.68 & 0.78 & 0.71 & 0.82 & 0.88 & 0.70 & 0.74 & 0.76 & 0.66 & 0.71 & 0.84 & 0.75 & 0.67 & 0.63\\
\bottomrule
\end{tabular}
\end{table}
\end{landscape}
\newpage

\begin{landscape}\begin{table}[!ht]

\caption{\label{tab:foo81}MASE from training set: 2006-2018 for yearly data}
\centering
\fontsize{8}{8}\selectfont
\begin{tabular}[t]{lrrrrrrrrrrrrrrrr}
\toprule
\multicolumn{1}{c}{ } & \multicolumn{4}{c}{Base Forecasting} & \multicolumn{6}{c}{HTS based on ARIMA} & \multicolumn{6}{c}{HTS based on ETS} \\
\cmidrule(l{3pt}r{3pt}){2-5} \cmidrule(l{3pt}r{3pt}){6-11} \cmidrule(l{3pt}r{3pt}){12-17}
series & arm & nve & ets & avg & bup & top & ols & mit & var & stc & ebup & etop & eols & emit & evar & estc\\
\midrule
\cellcolor{gray!6}{SL} & \cellcolor{gray!6}{2.15} & \cellcolor{gray!6}{1.55} & \cellcolor{gray!6}{1.24} & \cellcolor{gray!6}{2.15} & \cellcolor{gray!6}{1.31} & \cellcolor{gray!6}{2.15} & \cellcolor{gray!6}{2.12} & \cellcolor{gray!6}{1.51} & \cellcolor{gray!6}{1.55} & \cellcolor{gray!6}{1.84} & \cellcolor{gray!6}{1.48} & \cellcolor{gray!6}{1.24} & \cellcolor{gray!6}{1.27} & \cellcolor{gray!6}{1.44} & \cellcolor{gray!6}{1.44} & \cellcolor{gray!6}{1.38}\\
\addlinespace[0.3em]
\multicolumn{17}{l}{\textbf{Province}}\\
\hspace{1em}CN & 2.32 & 1.89 & 2.37 & 2.32 & 2.32 & 2.38 & 2.53 & 2.29 & 2.34 & 2.46 & 2.30 & 2.29 & 2.21 & 2.32 & 2.32 & 2.27\\
\hspace{1em}\cellcolor{gray!6}{ET} & \cellcolor{gray!6}{1.68} & \cellcolor{gray!6}{0.73} & \cellcolor{gray!6}{0.77} & \cellcolor{gray!6}{1.68} & \cellcolor{gray!6}{0.41} & \cellcolor{gray!6}{1.75} & \cellcolor{gray!6}{1.78} & \cellcolor{gray!6}{0.65} & \cellcolor{gray!6}{0.71} & \cellcolor{gray!6}{1.34} & \cellcolor{gray!6}{0.85} & \cellcolor{gray!6}{1.07} & \cellcolor{gray!6}{1.02} & \cellcolor{gray!6}{0.84} & \cellcolor{gray!6}{0.84} & \cellcolor{gray!6}{0.94}\\
\hspace{1em}NC & 0.83 & 0.36 & 0.42 & 0.83 & 0.83 & 0.89 & 1.67 & 0.79 & 0.83 & 1.25 & 0.39 & 0.30 & 0.15 & 0.40 & 0.40 & 0.22\\
\bottomrule
\end{tabular}
\end{table}
\end{landscape}

\newpage

\begin{landscape}\begin{table}[!ht]

\caption{\label{tab:foo}MASE for training set: 2016-2018 for weekly data}
\centering
\fontsize{8}{8}\selectfont
\begin{tabular}[t]{lrrrrrrrrrrrrrrrrrrr}
\toprule
\multicolumn{1}{c}{ } & \multicolumn{5}{c}{Base Forecasting} & \multicolumn{7}{c}{HTS based on ARIMA} & \multicolumn{7}{c}{HTS based on ETS} \\
\cmidrule(l{3pt}r{3pt}){2-6} \cmidrule(l{3pt}r{3pt}){7-13} \cmidrule(l{3pt}r{3pt}){14-20}
series & arm & snv & nve & ets & avg & bup & top & ols & mit & var & stc & cov & ebup & etop & eols & emit & evar & estc & ecov\\
\midrule
\cellcolor{gray!6}{SL} & \cellcolor{gray!6}{0.34} & \cellcolor{gray!6}{0.39} & \cellcolor{gray!6}{0.37} & \cellcolor{gray!6}{0.32} & \cellcolor{gray!6}{0.37} & \cellcolor{gray!6}{0.36} & \cellcolor{gray!6}{0.34} & \cellcolor{gray!6}{0.34} & \cellcolor{gray!6}{0.35} & \cellcolor{gray!6}{0.35} & \cellcolor{gray!6}{0.35} & \cellcolor{gray!6}{0.64} & \cellcolor{gray!6}{0.32} & \cellcolor{gray!6}{0.32} & \cellcolor{gray!6}{0.32} & \cellcolor{gray!6}{0.30} & \cellcolor{gray!6}{0.32} & \cellcolor{gray!6}{0.32} & \cellcolor{gray!6}{0.46}\\
\addlinespace[0.3em]
\multicolumn{20}{l}{\textbf{Province}}\\
\hspace{1em}CN & 0.53 & 0.57 & 0.60 & 0.57 & 0.59 & 0.53 & 0.53 & 0.53 & 0.53 & 0.53 & 0.53 & 0.76 & 0.51 & 0.58 & 0.57 & 0.50 & 0.53 & 0.53 & 1.00\\
\hspace{1em}\cellcolor{gray!6}{ET} & \cellcolor{gray!6}{0.38} & \cellcolor{gray!6}{0.64} & \cellcolor{gray!6}{0.38} & \cellcolor{gray!6}{0.60} & \cellcolor{gray!6}{0.54} & \cellcolor{gray!6}{0.39} & \cellcolor{gray!6}{0.39} & \cellcolor{gray!6}{0.40} & \cellcolor{gray!6}{0.37} & \cellcolor{gray!6}{0.39} & \cellcolor{gray!6}{0.40} & \cellcolor{gray!6}{0.72} & \cellcolor{gray!6}{0.38} & \cellcolor{gray!6}{0.55} & \cellcolor{gray!6}{0.51} & \cellcolor{gray!6}{0.44} & \cellcolor{gray!6}{0.46} & \cellcolor{gray!6}{0.47} & \cellcolor{gray!6}{0.38}\\
\hspace{1em}NC & 0.31 & 0.33 & 0.25 & 0.25 & 0.34 & 0.30 & 0.32 & 0.35 & 0.30 & 0.30 & 0.32 & 0.59 & 0.25 & 0.25 & 0.28 & 0.24 & 0.25 & 0.25 & 0.16\\
\bottomrule
\end{tabular}
\end{table}
\end{landscape}

\newpage

\begin{landscape}\begin{table}[!ht]

\caption{\label{tab:foo1}MASE from training set: 2016-2018 for monthly data}
\centering
\fontsize{8}{8}\selectfont
\begin{tabular}[t]{lrrrrrrrrrrrrrrrrrrr}
\toprule
\multicolumn{1}{c}{ } & \multicolumn{5}{c}{Base Forecasting} & \multicolumn{7}{c}{HTS based on ARIMA} & \multicolumn{7}{c}{HTS based on ETS} \\
\cmidrule(l{3pt}r{3pt}){2-6} \cmidrule(l{3pt}r{3pt}){7-13} \cmidrule(l{3pt}r{3pt}){14-20}
series & arm & snv & nve & ets & avg & bup & top & ols & mit & var & stc & cov & ebup & etop & eols & emit & evar & estc & ecov\\
\midrule
\cellcolor{gray!6}{SL} & \cellcolor{gray!6}{0.36} & \cellcolor{gray!6}{0.40} & \cellcolor{gray!6}{0.37} & \cellcolor{gray!6}{0.37} & \cellcolor{gray!6}{0.37} & \cellcolor{gray!6}{0.38} & \cellcolor{gray!6}{0.36} & \cellcolor{gray!6}{0.36} & \cellcolor{gray!6}{0.36} & \cellcolor{gray!6}{0.37} & \cellcolor{gray!6}{0.37} & \cellcolor{gray!6}{0.35} & \cellcolor{gray!6}{0.31} & \cellcolor{gray!6}{0.37} & \cellcolor{gray!6}{0.37} & \cellcolor{gray!6}{0.41} & \cellcolor{gray!6}{0.33} & \cellcolor{gray!6}{0.34} & \cellcolor{gray!6}{0.45}\\
\addlinespace[0.3em]
\multicolumn{20}{l}{\textbf{Province}}\\
\hspace{1em}CN & 0.56 & 0.58 & 0.58 & 0.56 & 0.60 & 0.56 & 0.56 & 0.56 & 0.55 & 0.56 & 0.56 & 0.54 & 0.53 & 0.59 & 0.61 & 0.64 & 0.54 & 0.57 & 0.67\\
\hspace{1em}\cellcolor{gray!6}{ET} & \cellcolor{gray!6}{0.52} & \cellcolor{gray!6}{0.67} & \cellcolor{gray!6}{0.42} & \cellcolor{gray!6}{0.59} & \cellcolor{gray!6}{0.54} & \cellcolor{gray!6}{0.45} & \cellcolor{gray!6}{0.52} & \cellcolor{gray!6}{0.52} & \cellcolor{gray!6}{0.49} & \cellcolor{gray!6}{0.49} & \cellcolor{gray!6}{0.50} & \cellcolor{gray!6}{0.52} & \cellcolor{gray!6}{0.54} & \cellcolor{gray!6}{0.50} & \cellcolor{gray!6}{0.49} & \cellcolor{gray!6}{0.56} & \cellcolor{gray!6}{0.56} & \cellcolor{gray!6}{0.52} & \cellcolor{gray!6}{0.66}\\
\hspace{1em}NC & 0.29 & 0.32 & 0.23 & 0.23 & 0.32 & 0.29 & 0.29 & 0.32 & 0.30 & 0.29 & 0.30 & 0.20 & 0.29 & 0.25 & 0.53 & 0.36 & 0.26 & 0.31 & 0.38\\
\bottomrule
\end{tabular}
\end{table}
\end{landscape}

\newpage

\begin{landscape}\begin{table}[!ht]

\caption{\label{tab:foo2}MASE from training set: 2016-2018 for quarterly data}
\centering
\fontsize{8}{8}\selectfont
\begin{tabular}[t]{lrrrrrrrrrrrrrrrrr}
\toprule
\multicolumn{1}{c}{ } & \multicolumn{5}{c}{Base Forecasting} & \multicolumn{6}{c}{HTS based on ARIMA} & \multicolumn{6}{c}{HTS based on ETS} \\
\cmidrule(l{3pt}r{3pt}){2-6} \cmidrule(l{3pt}r{3pt}){7-12} \cmidrule(l{3pt}r{3pt}){13-18}
series & arm & snv & nve & ets & avg & bup & top & ols & mit & var & stc & ebup & etop & eols & emit & evar & estc\\
\midrule
\cellcolor{gray!6}{SL} & \cellcolor{gray!6}{0.37} & \cellcolor{gray!6}{0.42} & \cellcolor{gray!6}{0.41} & \cellcolor{gray!6}{0.41} & \cellcolor{gray!6}{0.41} & \cellcolor{gray!6}{0.42} & \cellcolor{gray!6}{0.37} & \cellcolor{gray!6}{0.38} & \cellcolor{gray!6}{0.40} & \cellcolor{gray!6}{0.41} & \cellcolor{gray!6}{0.40} & \cellcolor{gray!6}{0.39} & \cellcolor{gray!6}{0.41} & \cellcolor{gray!6}{0.41} & \cellcolor{gray!6}{0.41} & \cellcolor{gray!6}{0.40} & \cellcolor{gray!6}{0.40}\\
\addlinespace[0.3em]
\multicolumn{18}{l}{\textbf{Province}}\\
\hspace{1em}CN & 0.59 & 0.62 & 0.63 & 0.65 & 0.63 & 0.59 & 0.55 & 0.56 & 0.59 & 0.59 & 0.58 & 0.54 & 0.69 & 0.66 & 0.65 & 0.58 & 0.60\\
\hspace{1em}\cellcolor{gray!6}{ET} & \cellcolor{gray!6}{0.56} & \cellcolor{gray!6}{0.77} & \cellcolor{gray!6}{0.40} & \cellcolor{gray!6}{1.95} & \cellcolor{gray!6}{0.56} & \cellcolor{gray!6}{0.77} & \cellcolor{gray!6}{0.59} & \cellcolor{gray!6}{0.64} & \cellcolor{gray!6}{0.69} & \cellcolor{gray!6}{0.71} & \cellcolor{gray!6}{0.69} & \cellcolor{gray!6}{0.48} & \cellcolor{gray!6}{1.34} & \cellcolor{gray!6}{1.42} & \cellcolor{gray!6}{0.73} & \cellcolor{gray!6}{0.74} & \cellcolor{gray!6}{0.92}\\
\hspace{1em}NC & 0.35 & 0.29 & 0.26 & 0.29 & 0.34 & 0.25 & 0.33 & 0.27 & 0.25 & 0.28 & 0.23 & 0.23 & 0.23 & 0.22 & 0.19 & 0.25 & 0.22\\
\bottomrule
\end{tabular}
\end{table}
\end{landscape}

\newpage
\bibliographystyle{new}
\bibliography{sample}

\end{document}